\documentclass{aastex63}
\usepackage{booktabs}

\received{2020 January 24}
\revised{2020 February 14}
\accepted{2020 February 14}
\submitjournal{ApJ}

\shorttitle{EUV and SXR Diffraction Efficiency of a Blazed Reflection Grating Fabricated by TASTE}
\shortauthors{McCoy et al.}
\watermark{}

\begin{document}

\title{Extreme Ultraviolet and Soft X-ray Diffraction Efficiency of a Blazed Reflection Grating Fabricated by Thermally Activated Selective Topography Equilibration}

\correspondingauthor{Jake McCoy}
\email{jam1117@psu.edu}

\author[0000-0002-1605-7517]{Jake A.\ McCoy}
\affiliation{Department of Astronomy \& Astrophysics, The Pennsylvania State University \\
525 Davey Laboratory \\
University Park, PA 16802, USA}

\author[0000-0003-2255-2968]{Randall L.\ McEntaffer}
\affiliation{Department of Astronomy \& Astrophysics, The Pennsylvania State University \\
525 Davey Laboratory \\
University Park, PA 16802, USA}

\author[0000-0001-5982-0060]{Drew M.\ Miles}
\affiliation{Department of Astronomy \& Astrophysics, The Pennsylvania State University \\
525 Davey Laboratory \\
University Park, PA 16802, USA}



\begin{abstract}
Future observatories utilizing reflection grating spectrometers for extreme ultraviolet (EUV) and soft x-ray (SXR) spectroscopy require high-fidelity gratings with both blazed groove facets and custom groove layouts that are often fanned or feature a slight curvature. 
While fabrication procedures centering on wet anisotropic etching in mono-crystalline silicon produce highly-efficient blazed gratings, the precision of a non-parallel groove layout is limited by the cubic structure of the silicon crystal. 
This motivates the pursuit of alternative techniques to grating manufacture, namely thermally activated selective topography equilibration (TASTE), which uses grayscale electron-beam lithography to pattern multi-leveled structures in resist followed by an optimized polymer thermal reflow to smooth the 3D patterns into continuous surface relief profiles. 
Using TASTE, a mold for a reflection grating with a periodicity of 400~nm and grooves resembling an asymmetric sawtooth was patterned in 130~nm-thick poly(methyl methacrylate) resist on a silicon substrate over a 50~mm by 7.5~mm area. 
This structure was coated with 15~nm of gold by electron-beam physical vapor deposition using titanium as an adhesion layer and then tested for EUV and SXR diffraction efficiency at beamline 6.3.2 of the Advanced Light Source synchrotron facility. 
Results demonstrate a quasi-blaze response characteristic of a $27^{\circ}$ blaze angle with groove facets smooth to 1.5~nm RMS. 
Absolute peak order efficiency ranges from 75\% to 25\% while total relative efficiency measures $\gtrapprox$90\% across the measured bandpass of 15.5~nm~$> \lambda >$~1.55~nm. 
\end{abstract}



\section{Introduction} \label{sec:introduction}
Instrumentation currently under development for spectroscopy at extreme ultraviolet (EUV) and soft x-ray (SXR) wavelengths ($\lambda \approx 40$ -- 0.5~nm) calls for blazed reflection gratings with sawtooth-shaped grooves and custom groove layouts to achieve both high spectral sensitivity and high spectral resolving power, $\lambda / \Delta \lambda$, in a grazing-incidence telescope. 
With a main scientific objective of measuring the diffuse, highly-ionized baryonic content in galactic halos and the intergalactic medium through SXR absorption spectroscopy of active galactic nuclei, the \emph{Lynx X-ray Observatory} is one of four flagship mission concepts considered for the 2020 Astrophysics Decadal Survey \citep{Gaskin19}. 
In a similar manner to the \emph{Reflection Grating Spectrometer (RGS)} on board \emph{XMM-Newton} \citep{denHerder01}, an x-ray reflection grating spectrometer suitable for \emph{Lynx} requires several thousand identical blazed gratings with fanned groove layouts stacked and aligned into modular arrays to intercept SXR radiation coming to a focus in a Wolter-I telescope \citep{McEntaffer19}. 
On the other hand, the \emph{Extreme-Ultraviolet Stellar Characterization for Atmospheric Physics and Evolution (ESCAPE)} mission concept incorporates two blazed gratings with curved groove layouts that play a similar role for EUV radiation in a Hettrick-Bowyer-I telescope with the goal of characterizing high-energy radiation in habitable zones surrounding M-dwarfs and their impact on the atmospheres of exoplanets \citep{France19}. 
A main challenge from the standpoint of grating fabrication in any case is the realization of a lithographic process that can generate non-parallel groove layouts with high fidelity while also maintaining blazed grooves that enable high diffraction efficiency. 
In particular, sensitivity requirements for \emph{Lynx} require that the sum of all propagating orders exceeds 40\% diffraction efficiency across the SXR bandpass while \emph{ESCAPE} baselines single-order diffraction efficiency of $\sim$60\% in the EUV \citep{McEntaffer19,France19}. 

The state-of-the-art for blazed gratings that perform with high diffraction efficiency at EUV and SXR wavelengths are those fabricated by wet anisotropic etching in mono-crystalline silicon, where typically either interference lithography \citep{Franke97,Chang03} or electron-beam lithography \citep{Voronov11,Miles18} is used to define a groove layout in resist before the pattern is transferred into the underlying silicon crystal structure to produce atomically-smooth sawtooth facets. 
However, interference lithography faces severe limitations in its ability to pattern non-parallel layouts and even with the direct-write capabilities of electron-beam lithography, the cubic structure of mono-crystalline silicon prevents the formation of fanned or curved grooves with smooth and continuous triangular facets. 
Additionally, these anisotropic etching processes demand precise alignment between the groove layout in resist and the crystallographic planes of silicon to produce a high-fidelity grating. 
An alternative to these methods of grating manufacture is thermally activated selective topography equilibration (TASTE), which combines grayscale electron-beam lithography (GEBL) and polymer thermal reflow to produce smooth, 3D surface relief profiles in poly(methyl methacrylate) (PMMA) or other thermoplastic resists such as ZEP520A and mr-PosEBR \citep{Schleunitz14,Kirchner16,Pfirrmann16}. 
Through optimization of TASTE, repeating staircase patterns in PMMA fabricated by GEBL can be equilibrated into wedgelike structures by selective thermal reflow to provide a template for a blazed grating with groove spacing on the order of hundreds of nanometers \citep{McCoy18}. 
With no dependence on the crystallographic structure of the substrate, TASTE has the potential for realizing reflection gratings that feature both a blazed surface topography and a non-parallel groove layout, thereby enabling high sensitivity and high $\lambda / \Delta \lambda$ in an EUV/SXR spectrometer.  

This paper presents diffraction efficiency measurements of a grating prototype fabricated using TASTE that emulates a blazed grating with a uniform groove spacing of 400~nm and a blaze angle of $\sim$27$^{\circ}$. 
Gathered at beamline 6.3.2 for EUV and SXR reflectometry of the Advanced Light Source (ALS) synchrotron facility at Lawrence Berkeley National Laboratory\footnote{\url{http://cxro.lbl.gov/als632/}} \citep{Underwood96,Gullikson01}, these measurements characterize the efficiency response of the grating in an extreme off-plane mount at a graze angle of $\sim$1.5$^{\circ}$ to enable total external reflection at SXR wavelengths. 
These results serve as the first demonstration of TASTE being used for EUV/SXR grating technology and provide a baseline for further experimentation with gratings that feature non-parallel grooves. 
Both the beamline test campaign and the grating prototype fabrication procedure are described in Section~\ref{sec:experiment}, with all processing for grating fabrication and materials characterization carried out at the Pennsylvania State University (PSU) Materials Research Institute.\footnote{\url{https://www.mri.psu.edu}} 
The beamline measurements are presented in Section~\ref{sec:results} and discussed in Section~\ref{sec:discussion} before conclusions and a summary are provided in Section~\ref{sec:summary}. 

\section{Experiment and Grating Fabrication}\label{sec:experiment}
This section introduces the beamline test procedure used to measure EUV and SXR diffraction efficiency and details how the grating prototype was fabricated. 
Based on the geometrical considerations for this test campaign outlined in Section~\ref{sec:als_testing}, the grating grooves were patterned in PMMA on a silicon wafer using the TASTE procedure described in Section~\ref{sec:taste_procedure} and then coated with gold for EUV and SXR reflectivity as discussed in Section~\ref{sec:coating}. 

\subsection{Diffraction Efficiency Testing at the Advanced Light Source (ALS)}\label{sec:als_testing}
\begin{figure}
 \centering
 \includegraphics[scale=0.4]{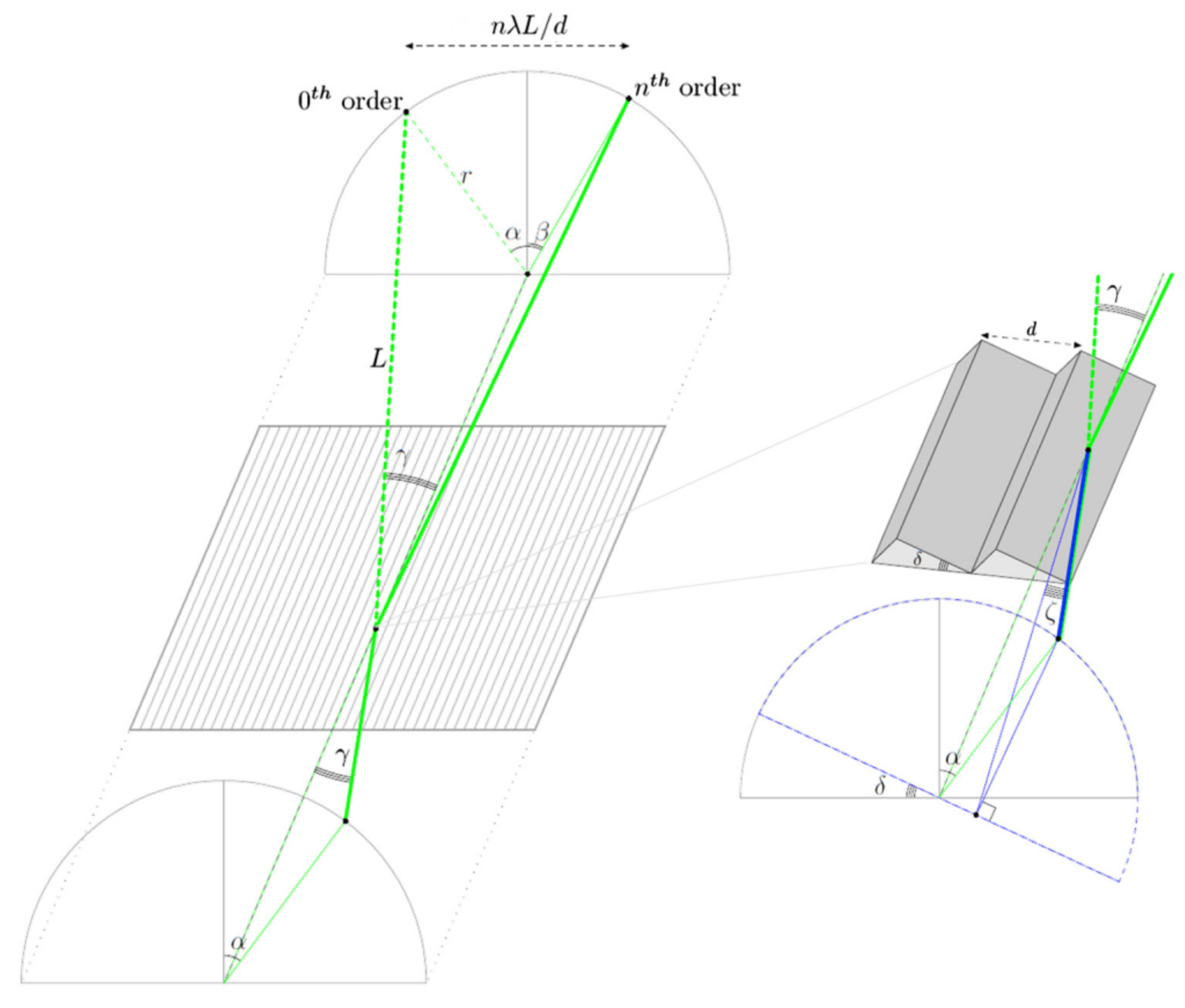}
 \caption{Geometry for a reflection grating producing a conical diffraction pattern. In an extreme off-plane mount, the incoming radiation is nearly parallel to the groove direction where the half-angle of the cone opening, $\gamma$, is on the order of a degree while the azimuthal incidence angle, $\alpha$, can take on any value to match the blaze angle, $\delta$, in Littrow configuration while also maintaining an incidence angle on the groove facets, $\zeta$, that is smaller than the critical angle for total external reflection at extreme ultraviolet and soft x-ray wavelengths. At a distance $L$ away from the point of incidence on the grating, diffracted orders are each separated by a distance $\lambda L / d$ along the direction of grating periodicity (\emph{i.e.}, the dispersion direction), where $d$ is the groove spacing. Figure taken from \citet{McCoy18}.}\label{fig:conical_reflection}
 \end{figure}

Beamline 6.3.2 of the ALS provides a station for EUV and SXR reflectometry where a highly coherent, tunable beam of monochromatic radiation with wavelength 40~nm~$\gtrapprox \lambda \gtrapprox$~1~nm under high vacuum is incident onto a stage-mounted optic while a photodiode detector is used to measure the intensity of outgoing radiation. 
Using this laboratory facility, absolute diffraction efficiency of a grating as a function of $\lambda$, defined as the intensity ratio between the $n^{\text{th}}$ diffracted order and the unobstructed beam
\begin{equation}\label{eq:diffraction_efficiency}
 \mathcal{E}_n \! \left( \lambda \right)  \equiv \frac{\mathcal{I}_n \! \left( \lambda \right)}{\mathcal{I}_{\text{inc}} \! \left( \lambda \right)} ,
 \end{equation}
can be determined experimentally following the test procedures outlined by \citet{Miles18}, where the photodiode detector mounted on vertical goniometric and horizontal linear staging at a distance $L \approx 235$~mm away from the point of incidence on the grating is used to measure $\mathcal{I}_n \! \left( \lambda \right)$ for each propagating order and $\mathcal{I}_{\text{inc}} \! \left( \lambda \right)$. 
The grating prototype described in this paper was designed specifically for taking diffraction efficiency measurements at this beamline in a grazing-incidence, extreme off-plane mount where the incident radiation is nearly parallel to the groove direction and propagating orders are confined to the surface of a cone with a small opening angle \citep{Cash91}. 
The locations of orders for radiation of wavelength $\lambda$ diffracting from a grating with a groove spacing $d$ are described by the generalized grating equation \citep{Neviere78}
\begin{equation}\label{eq:off-plane_incidence_orders}
 \sin \left( \alpha \right) + \sin \left( \beta \right) = \frac{n \lambda}{d \sin \left( \gamma \right)} \quad \text{for } n = 0, \pm 1, \pm 2, \pm 3 ... 
 \end{equation}
where, as illustrated in the left panel of Figure~\ref{fig:conical_reflection}, $\gamma$ is the half-opening angle of the cone, $\alpha$ is the azimuthal incidence angle and $\beta$ is the azimuthal diffracted angle of the $n^{\text{th}}$ diffracted order. 
The testing methodology adopted from \citet{Miles18} relies on the radius of the diffracted arc, given by 
\begin{equation}\label{eq:arc_radius}
 r = L \sin \left( \gamma \right) ,
 \end{equation}
being smaller than the 10~mm by 10~mm collecting area of the photodiode detector used at the beamline. 
With propagating orders being dispersed a distance from $0^{\text{th}}$ order along the cross-groove direction given by
\begin{equation}\label{eq:linear_dispersion}
 x_n = \frac{n \lambda L}{d} 
 \end{equation}
as shown in in Figure~\ref{fig:conical_reflection}, which is in this case on the order of millimeters, a vertical, 0.5~mm-wide slit masking the detector is installed to enable the intensity of each diffracted order, $\mathcal{I}_n \! \left( \lambda \right)$, to be measured in isolation as the diffracted arc is scanned along the direction of the horizontal linear staging. 
Moreover, $\mathcal{I}_{\text{inc}} \! \left( \lambda \right)$ is measured in a similar fashion when the grating is moved out of the path of the beam using controllable staging. 
In this way, diffraction efficiency according to equation~\ref{eq:diffraction_efficiency} can be measured at EUV and SXR wavelengths by repeating this process for many values of $\lambda$ using the tunable beam provided by the ALS. 

\begin{figure}
 \centering
 \includegraphics[scale=0.4]{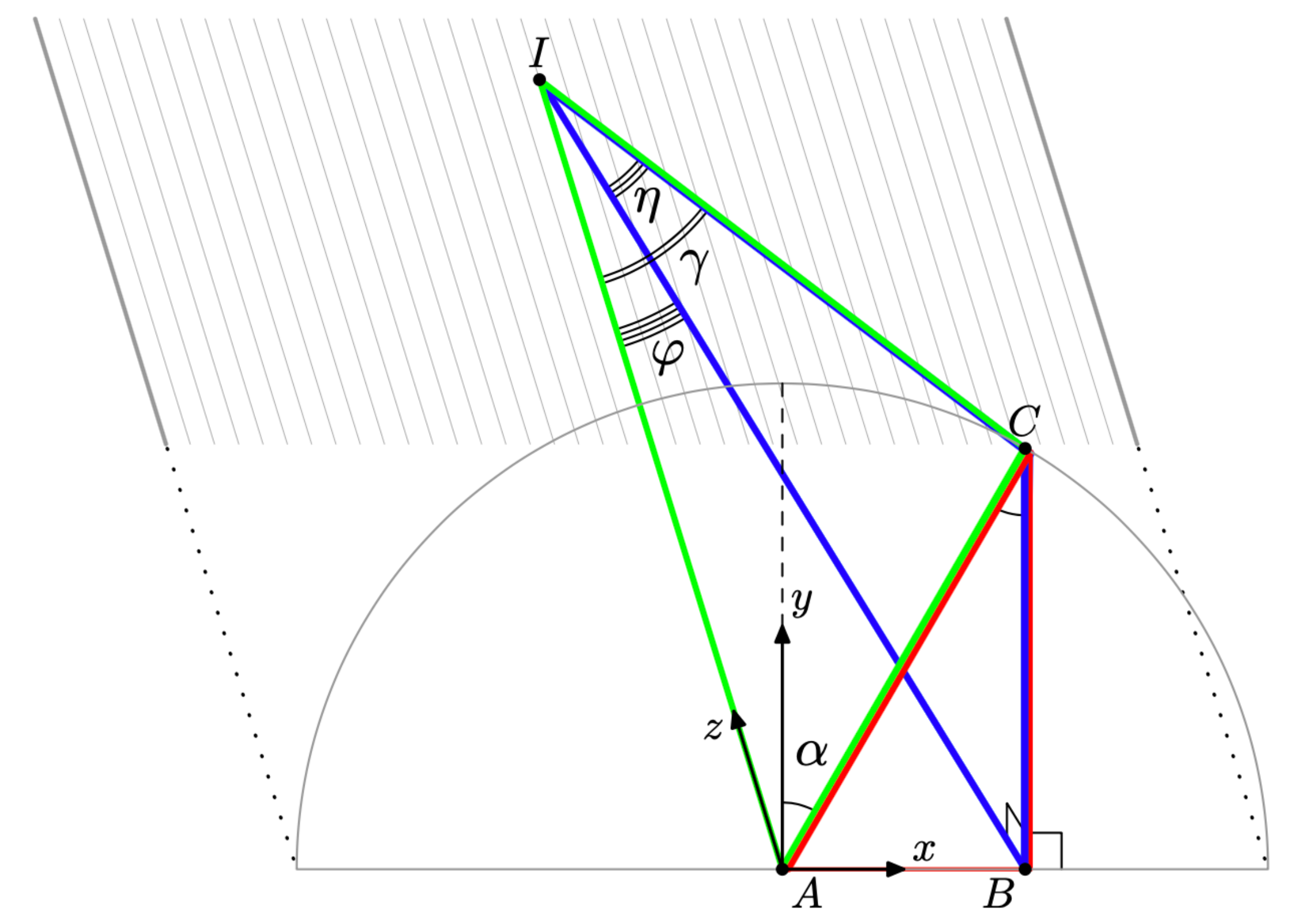}
 \caption{Angles relevant for beamline diffraction efficiency testing. Grating incidence angles $\alpha \equiv \angle ACB$ and $\gamma \equiv \angle AIC$ illustrated in Figure~\ref{fig:conical_reflection} are controlled through the adjustment of principal axis angles $\eta \equiv \angle CIB$ and $\varphi \equiv \angle AIB$ at beamline 6.3.2 of the Advanced Light Source.}\label{fig:grating_angles} 
 \end{figure}

As described in Section~\ref{sec:taste_procedure}, the groove spacing of the grating prototype was designed to be $d=400$~nm while the angle of the sawtooth facets achieved by TASTE in 130~nm-thick PMMA yields a blaze angle of $\delta \sim 27^{\circ}$. 
To enable an effective blaze response from the grating so that $\mathcal{E}_n \! \left( \lambda \right)$ is concentrated in a particular part of the spectrum, $\alpha$ and $\gamma$ should be set such that only the shallow side of the asymmetric, sawtooth-shaped grooves is illuminated \citep{Loewen97}. 
Radiation is incident on these sawtooth facets at an angle $\zeta$ as illustrated in the right panel of Figure~\ref{fig:conical_reflection}, which must be smaller than the critical angle for total external reflection \citep{Attwood17} and is related to $\alpha$ and $\gamma$ through the following relation:
\begin{equation}\label{eq:angle_on_groove}
 \sin \left( \zeta \right) = \sin \left( \gamma \right) \cos \left( \delta - \alpha \right) .
 \end{equation} 
The result in principle is that radiation is preferentially diffracted to an angle $\beta = 2 \delta - \alpha$ so that the blaze wavelength for the $n^{\text{th}}$ propagating order is 
\begin{equation}\label{eq:blaze_wavelength}
 \lambda_b = \frac{d \sin \left( \gamma \right)}{n} \left[ \sin \left( \alpha \right) + \sin \left( 2 \delta - \alpha \right)  \right] .
 \end{equation}
At the beamline, $\alpha$ and $\gamma$ are controlled through the movement of stage rotations along principal axes relative to the surface of the grating substrate. 
Referencing the Cartesian coordinate system drawn in Figure~\ref{fig:grating_angles}, the stage-controllable angles are rotations about the $x$ and $y$ axes. 
Rotations about the $x$ axis at the point of incidence are tied to the graze angle relative to the surface of the grating, $\eta$, defined by the relation 
\begin{equation}\label{eq:graze_angle}
 \sin \left( \eta \right) = \sin \left( \gamma \right) \cos \left( \alpha \right) ,
 \end{equation}
whereas rotations about the $y$ axis represent grating yaw, $\varphi$, which is related to $\alpha$ and $\eta$ by 
\begin{equation}\label{eq:yaw_angle}
 \sin \left( \varphi \right) = \tan \left( \alpha \right) \tan \left( \eta \right) ,
 \end{equation}
where $\varphi = 90^{\circ}$ corresponds to an exact in-plane mount with $\sin \left( \gamma \right) = 1$. 
Additionally, the orientation of the grating relative to the $z$ axis is characterized by the roll angle, $\phi$, which serves to rotate order locations about the center of the diffracted arc. 
This angle, not shown in Figure~\ref{fig:grating_angles}, remains fixed nominally at $\phi = 0^{\circ}$ but like $\eta$ and $\varphi$, it must be constrained to measure $\alpha$ and $\gamma$ accurately; this is addressed in Section~\ref{sec:results}. 

To satisfy the blaze condition for the grating prototype and the testing methodology requirement that $\gamma \lessapprox 2^{\circ}$ comfortably, the grating prototype was designed for use at a nominal graze angle of $\eta = 1.5^{\circ}$ in a Littrow configuration, where $\alpha = \beta = \delta \approx 27^{\circ}$ and $\zeta = \gamma \approx 1.7^{\circ}$ by equations \ref{eq:angle_on_groove} and \ref{eq:graze_angle}. 
In this configuration, diffraction efficiency is expected to be maximized and equation~\ref{eq:blaze_wavelength} for the blaze wavelength becomes 
\begin{equation}\label{eq:blaze_wavelength_littrow}
 \lambda_b = \frac{2 d \sin \left( \gamma \right) \sin \left( \delta \right) }{n} \approx \frac{11~\text{nm}}{n}.
 \end{equation}
Meanwhile, equation~\ref{eq:yaw_angle} yields $\varphi \approx 0.8^{\circ}$ so that for $\phi \approx 0^{\circ}$, the grating dispersion direction is virtually parallel to the direction of the horizontal linear stage motion. 
Because the incident beam is then nearly parallel with both the groove direction and the surface of the grating substrate, the grooves of the grating prototype must be long enough to encompass the incident beam in projection at the chosen grazing-incidence angle of $\eta = 1.5^{\circ}$. 
With knowledge that the cross-sectional size of the beam at the ALS is $\lessapprox 0.5$~mm as it is incident on an optic, the grating prototype was designed to be 50~mm along the groove direction and 7.5~mm along the dispersion direction as to allow the beam to be positioned on the grooved area with relative ease. 
Considering the EUV/SXR wavelengths at which there exist propagating orders in this geometry and the separation of these orders defined by equation~\ref{eq:linear_dispersion} with $d=400$~nm and $L \approx 235$~mm relative to the 0.5~mm slit width, diffraction efficiency testing at the ALS was restricted to 15.5~nm~$> \lambda >$~1.55~nm, or equivalently, photon energies ranging from 80~eV to 800~eV. 

\subsection{Thermally Activated Selective Topography Equilibration (TASTE)}\label{sec:taste_procedure}
\begin{figure}
 \centering
 \includegraphics[scale=0.35]{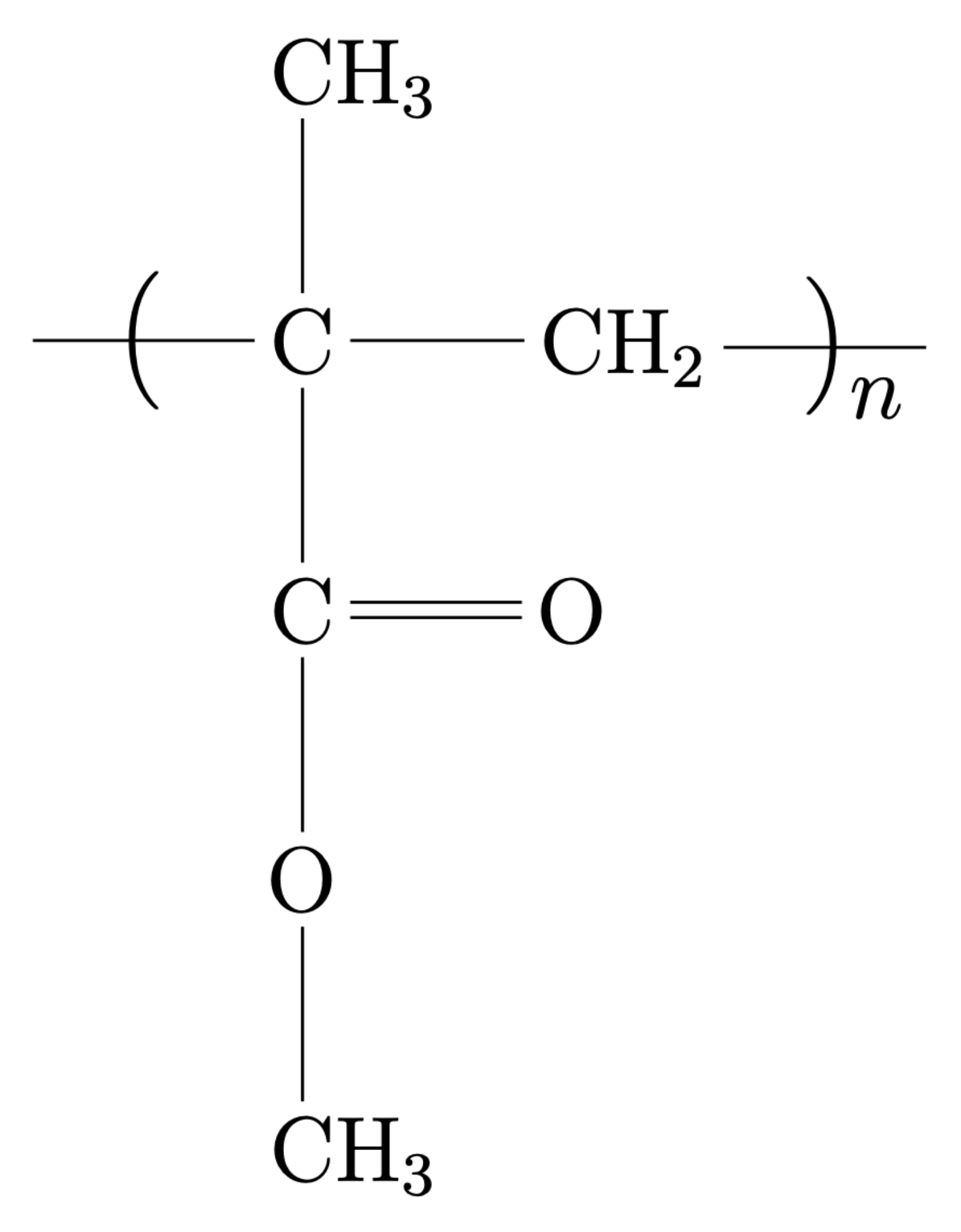}
 \caption{Structural chemical composition of poly(methyl methacrylate) (PMMA). The average molecular weight ($M_w$) of PMMA resist depends directly on the typical length of these polymer chains, which is indicated by the degree of polymerization, $n$ (\emph{i.e.}, the number of MMA monomers bonded together). }\label{fig:PMMA} 
 \end{figure}

The surface relief mold for the grating prototype was fabricated by TASTE in $130$~nm-thick PMMA coated on a silicon wafer at the Nanofabrication Laboratory of the PSU Materials Research Institute.\footnote{\url{https://www.mri.psu.edu/nanofabrication-lab}} 
As described by \citet{Schleunitz14}, TASTE consists of two main processes: grayscale electron-beam lithography (GEBL) to pattern multi-level structures in a thermoplastic resist such as PMMA and selective thermal reflow to equilibrate the topography into smooth, sloped surfaces. 
Both of these components depend on local modification of the average molecular weight in the resist, $M_w$, by lithographic exposure to high-energy electrons. 
The structural formula for PMMA is drawn in Figure~\ref{fig:PMMA}, where methyl methacrylate (MMA) monomers (C$_5$O$_2$H$_8$) are bonded together at the sites marked by brackets to form long polymer chains that constitute the resist as an amorphous material. 
The quantity $M_w$ is therefore dependent on the typical length of these polymer chains; more precisely, it is defined as the weight-averaged molar mass of the PMMA molecules making up the resist. 
A local reduction in $M_w$ occurs in positive-tone resists such as PMMA when high-energy electron exposure induces polymer chain scission by breaking bonds between monomers \citep{Dobisz00}. 
For the processing described in this paper, the 130~nm-thick resist film was attained by spin-coating PMMA with $M_w = 950$~kg~mol$^{-1}$ diluted 3\% in anisole (\textsc{MicroChem Corp.}) on a clean, dehydrated, polished silicon wafer 100~mm in diameter and 0.5~mm thick (\textsc{Virginia Semiconductor, Inc.}) at 3~krpm using a dynamic dispense followed by a solvent bake-out.\footnote{Both the dehydration bake and the solvent bake were performed at $180^{\circ}$C for 3 minutes by hotplate.}  
Lithographic electron-beam exposure, quantified as electron dose, $D$, was carried out using a \textsc{Raith EBPG5200} system\footnote{\url{https://www.raith.com/products/ebpg5200.html}} with a 100~kV accelerating voltage at the PSU Nanofabrication Laboratory. 

In standard electron-beam lithography, resist is exposed with a fixed, sufficiently large dose $D$ and therefore $M_w$ is locally reduced to a high degree. 
With a 100~kV accelerating voltage, electrons are energetic enough to forward-scatter through the resist with negligible intensity loss over a 130~nm thickness and therefore $M_w$ can be considered to be uniform throughout the depth of the resist film. 
This causes exposed resist to be soluble for wet development so that it can be etched down to the substrate while unexposed resist remains virtually intact. 
In constrast to this process, which produces a bi-level topography in resist, GEBL relies on a lateral gradient of $M_w$ imparted in the resist to produce a multi-level topography following a timed wet development \citep{Stauffer92}. 
\begin{figure}
 \centering
 \includegraphics[scale=0.5]{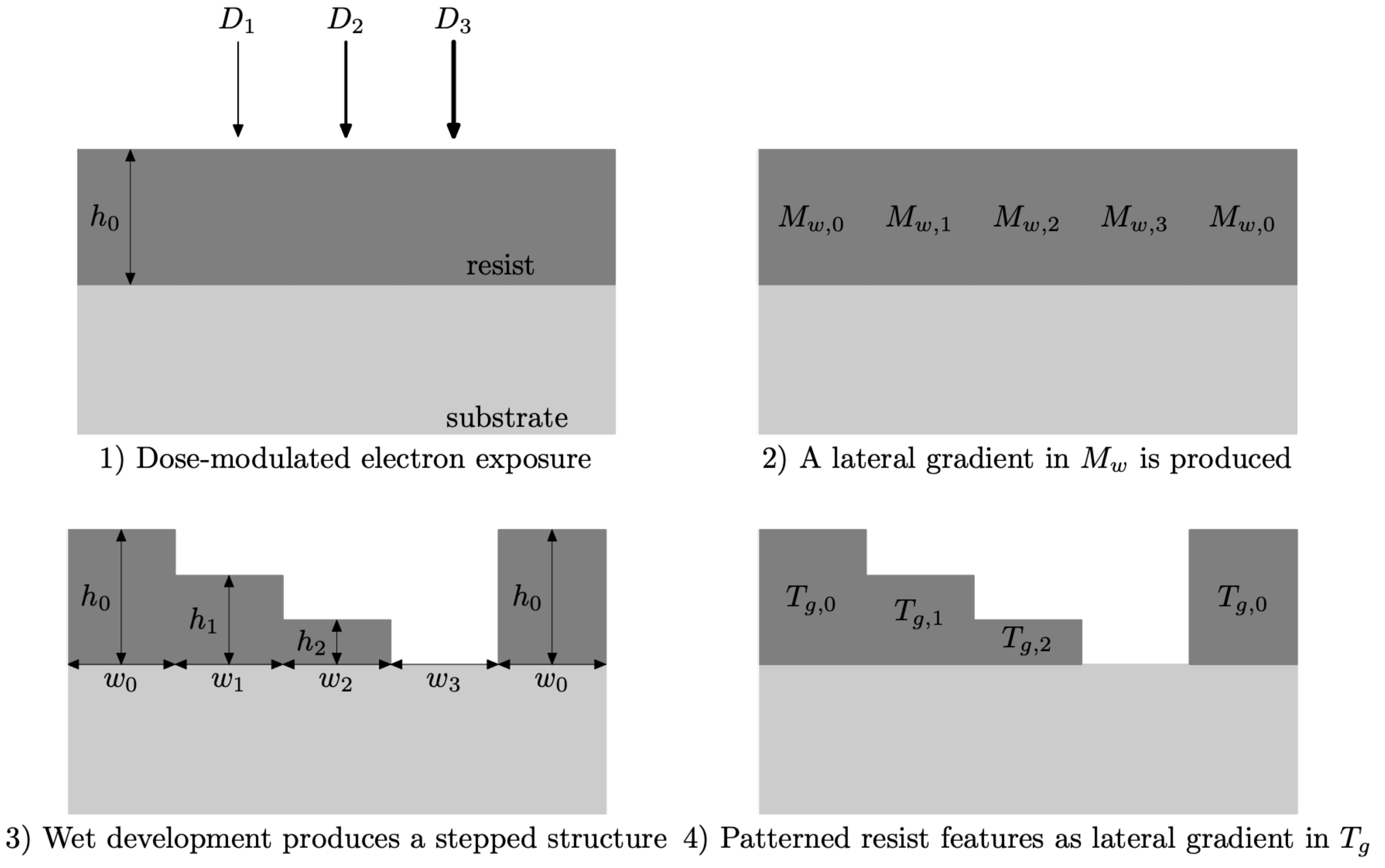}
 \caption{Physical properties of PMMA resist processed by grayscale electron-beam lithography (GEBL) that enable thermally activated selective topography equilibration. Dose-modulated electron exposure gives rise to a lateral gradient in average molecular weight ($M_w$) so that varying resist thicknesses result from $M_w$-dependent etch rates that occur during wet development. The GEBL-fabricated structure exhibits a lateral gradient in glass-liquid transition temperature ($T_g$), enabling selective thermal reflow. These illustrations neglect the effect of lateral development, which gives rise to tilted surfaces and rounded corners on the staircase steps.}\label{fig:TASTE_diagram}
 \end{figure}
Illustrated in Figure~\ref{fig:TASTE_diagram}, this is achieved using a dose-modulated electron exposure where doses $D_1 < D_2 < D_3$ give rise to local average molecular weights $M_{w,1} > M_{w,2} > M_{w,3}$, and if $D_3$ is large enough to clear the resist, local resist thicknesses $h_1 > h_2 > h_3 = 0$ following wet development.  
Unexposed portions of the resist (\emph{i.e.}, top steps of the staircase topography) in principle retain the original molecular weight of the polymer, $M_{w,0}$, and the coated resist thickness, $h_0$. 
While these portions of the resist may be inadvertently dosed by the proximity effect of electrons back-scattering through the substrate \citep{Pavkovich86}, the GEBL principles discussed here hold provided that there is sufficient contrast in $M_w$ following dose-modulated electron exposure and that an appropriate wet development recipe is adopted. 
Also dependent on $M_w$ is the polymer glass-liquid transition temperature, $T_g$, such that local average molecular weights $M_{w,0} > M_{w,1} > M_{w,2}$ correspond to $T_{g,0} > T_{g,1} > T_{g,2}$, where $T_{g,0}$ is the transition temperature of unexposed resist. 
Owing to the thermoplastic nature of the resist, the stepped structure produced by GEBL can be heated globally to a temperature $T_{\text{reflow}}$ such that electron-exposed resist is allowed to equilibrate in a molten state according to a time-dependent visco-elastic creep process while unexposed resist remains in its glass state \citep{Schleunitz10,Kirchner14}. 
In this way, selective thermal reflow can be achieved through heating the substrate by hotplate to a temperature $T_{g,0} > T_{\text{reflow}} > T_{g,1}$ and a stepped topography can be equilibrated into a sloped, sawtooth-like topography to serve as a surface relief mold for a blazed grating through optimization of GEBL parameters, $T_{\text{reflow}}$ and heating time. 

\begin{figure}
 \centering
 \includegraphics[scale=0.65]{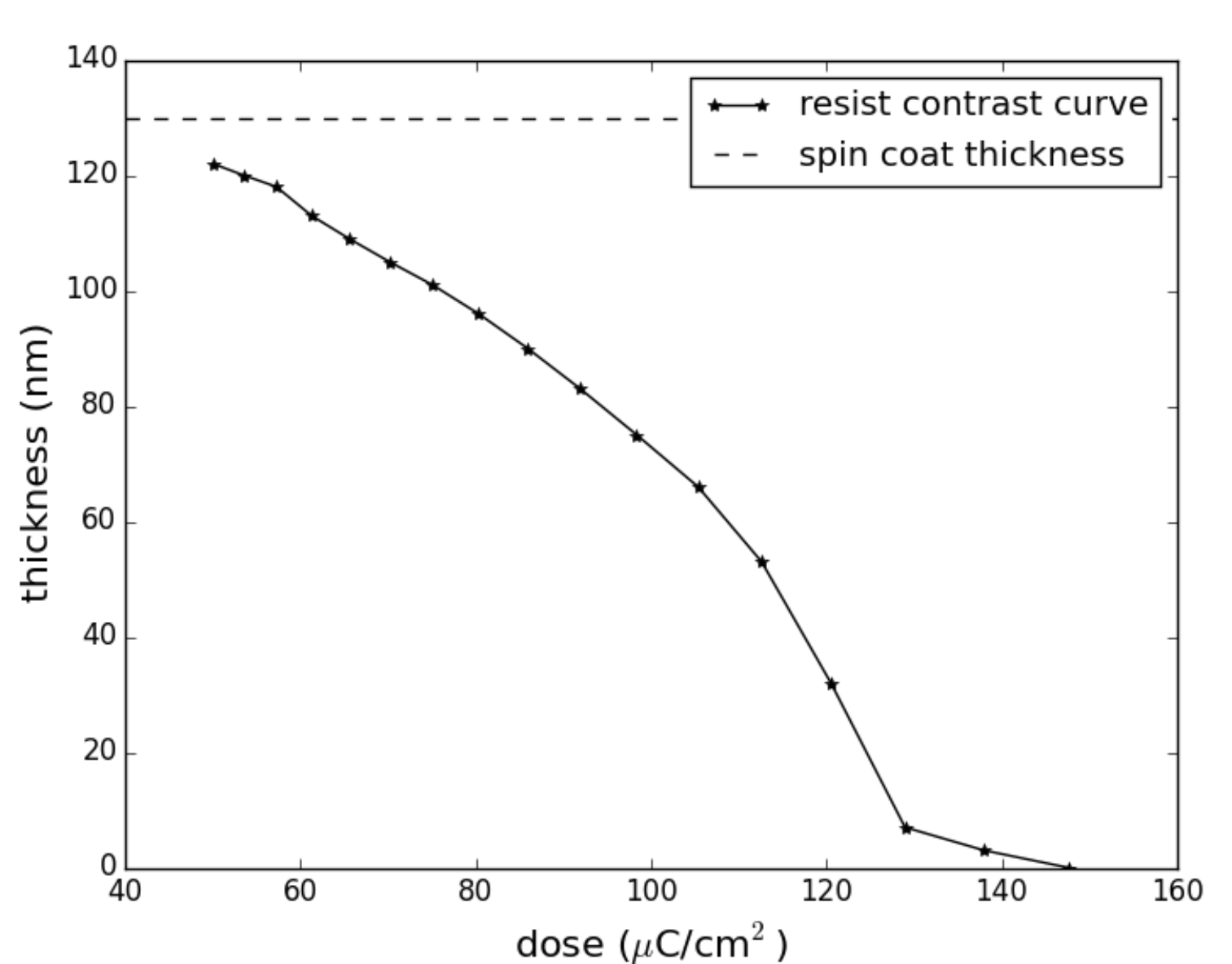}
 \caption{Resist contrast data for 130~nm-thick, 950~kg~mol$^{-1}$ PMMA developed at room temperature using 1:1 methyl isobutyl ketone (MIBK) and isopropyl alcohol (IPA)  for 2 minutes followed by a 30-second IPA rinse and a nitrogen blow dry. Figure taken from \citet{McCoy18}.}\label{fig:resist_contrast}
 \end{figure}

GEBL processing for fabrication of the grating prototype surface relief mold is outlined in Figure~\ref{fig:TASTE_diagram}: the staircase topography features two electron-exposed steps, a cleared area and an unexposed step, all of equal width consistent with a periodicity of $d = 400$~nm (\emph{i.e.}, $w_0 = w_1 = w_2 = w_3 = 100$~nm). 
Electron dosing for GEBL was performed according to the resist contrast curve provided by \citet{McCoy18}, which is based on a room-temperature development recipe consisting of 2 minutes in a 1:1 mixture of methyl isobutyl ketone (MIBK) and isopropyl alcohol (IPA) followed by a 30-second rinse in IPA and a high-purity nitrogen blow-dry. 
This contrast curve is shown in Figure~\ref{fig:resist_contrast}, where post-development PMMA thickness as measured by spectroscopic ellipsometry is plotted as a function of electron dose, $D$. 
These data were processed using the three-dimensional proximity effect correction (3DPEC) algorithm included in the \textsc{Layout Beamer} software package developed by \textsc{GenISys GmbH}\footnote{\url{https://genisys-gmbh.com/beamer.html}} \citep{Unal10} to generate a dose-corrected layout appropriate for achieving exposed staircase steps with $h_1 \approx 0.66 h_0$ and $h_2 \approx 0.33 h_0$, where $h_0 \approx 130$~nm is the spin-coat thickness. 
Electron exposure for GEBL was carried out using an 8~nA beam current and a 400~$\micron$ aperture with a beam step size and a writing grid resolution of 10~nm, which is comparable to the beam spot size realized by the \textsc{EBPG5200} under these conditions. 
These beam conditions differ from the recipe described in \citet{McCoy18}, which was limited by the 25~MHz EBPG5200 clock frequency at the time of publication. 

Using the GEBL process outlined above, test patterns were exposed, developed and characterized by atomic force microscopy (AFM) to verify that the previously-reported staircase topography could be readily reproduced using the increased value for beam current enabled by a clock frequency upgrade to the \textsc{EBPG5200}. 
\begin{figure}
 \centering
 \includegraphics[scale=0.7]{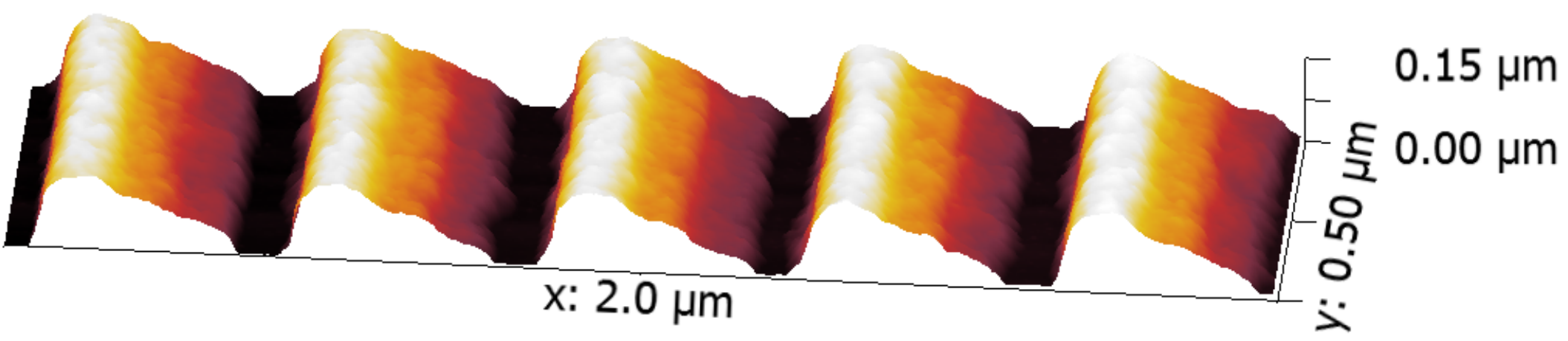}
 \includegraphics[scale=0.7]{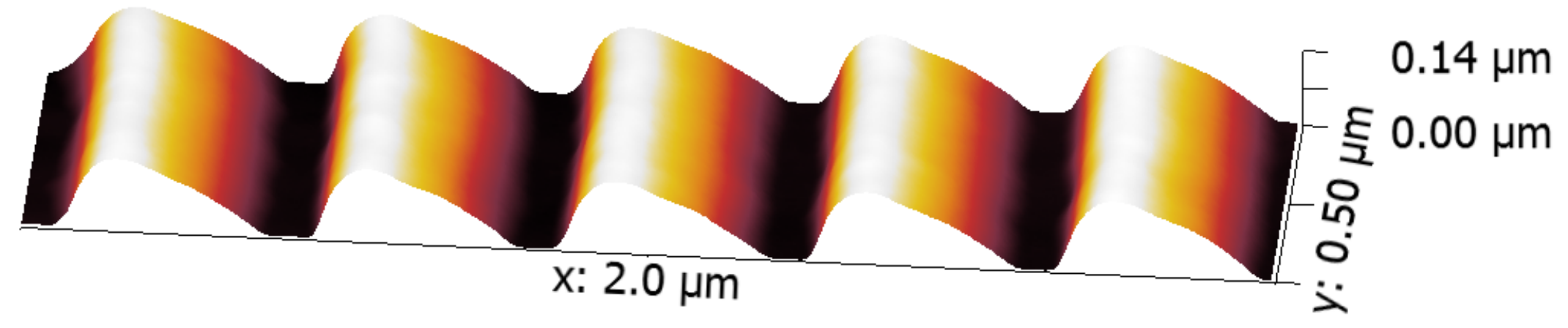}
 \caption{Atomic force micrographs of the GEBL-processed resist (\emph{top}) and the resist following thermal reflow (\emph{bottom}).}\label{fig:TASTE_AFMs}
 \end{figure}
In an identical fashion to \citet{McCoy18}, all AFM was carried out at the PSU Materials Characterization Laboratory\footnote{\url{https://www.mri.psu.edu/materials-characterization-lab}} with a \textsc{Bruker Icon} instrument equipped with a \textsc{SCANASYST-AIR} tip over 2~$\micron$ in the direction of grating periodicity at 512 samples per line to yield a 3.9~nm pixel size using \textsc{Bruker}'s PeakForce Tapping$^{\text{TM}}$ mode. 
A scan of the GEBL pattern exposed using an 8~nA beam current and a 400~$\micron$ aperture is shown in the top panel of Figure~\ref{fig:TASTE_AFMs}, where it is verified that the topography appears virtually indistinguishable from the previous result obtained using a 1~nA beam current and a 200~$\micron$ aperture. 
Next, thermal reflow experimentation on test samples was carried out using an automated hotplate tool on a resist stabilization system built by \textsc{Fusion Semiconductor}, where from the results reported by \citet{McCoy18}, it is expected that the optimum value for $T_{\text{reflow}}$ is near 120$^{\circ}$C. 
Through a series of reflow tests, it was found that $T_{\text{reflow}} = 116^{\circ}$C applied for a duration of 30 minutes\footnote{Due to the 999-second time-out of the \textsc{Fusion Semiconductor} automated hotplate tool, thermal reflow was carried out in two consecutive 15-minute intervals.} produced a topography that most closely resembled a sawtooth; an AFM of a test pattern treated this way is shown in the bottom panel of Figure~\ref{fig:TASTE_AFMs}. 
Based on these results, a 7.5~mm by 50~mm area was exposed for GEBL using a 300~$\micron$ by 300~$\micron$ mainfield with 10-nm resolution, a 4~$\micron$ by 4~$\micron$ subfield with 5-nm resolution and Large Rectangle Fine Trapezoid (LRFT) fracturing in \textsc{Layout Beamer}. 
\begin{figure}
 \centering
 \includegraphics[scale=0.12]{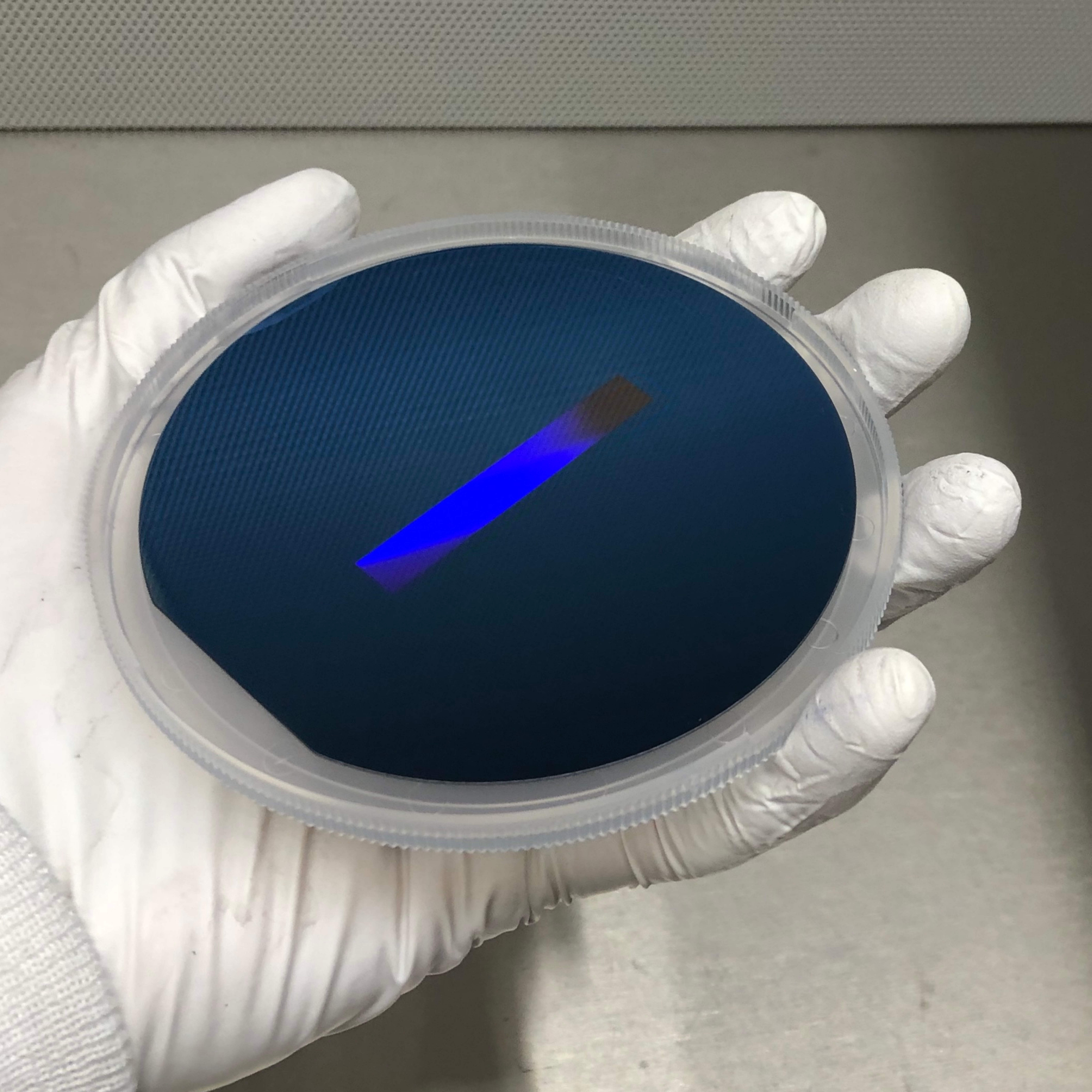}
 \caption{Surface relief mold for the grating prototype patterned in 130~nm-thick PMMA coated on a 4-inch silicon wafer using TASTE. The grating measures 50~mm in the groove direction and 7.5~m in the dispersion direction.}\label{fig:uncoated_grating_pic}
 \end{figure}
Under these conditions, the \textsc{EBPG5200} exposure duration (including tool overhead) was $\lessapprox 20$ hours. 
The grating mold resulting from the entire TASTE process patterned on a 4-inch wafer is pictured in Figure~\ref{fig:uncoated_grating_pic}. 

\subsection{Coating for Extreme Ultraviolet (EUV) and Soft X-ray (SXR) Reflectivity}\label{sec:coating}
\begin{figure}
 \centering
 \includegraphics[scale=0.55]{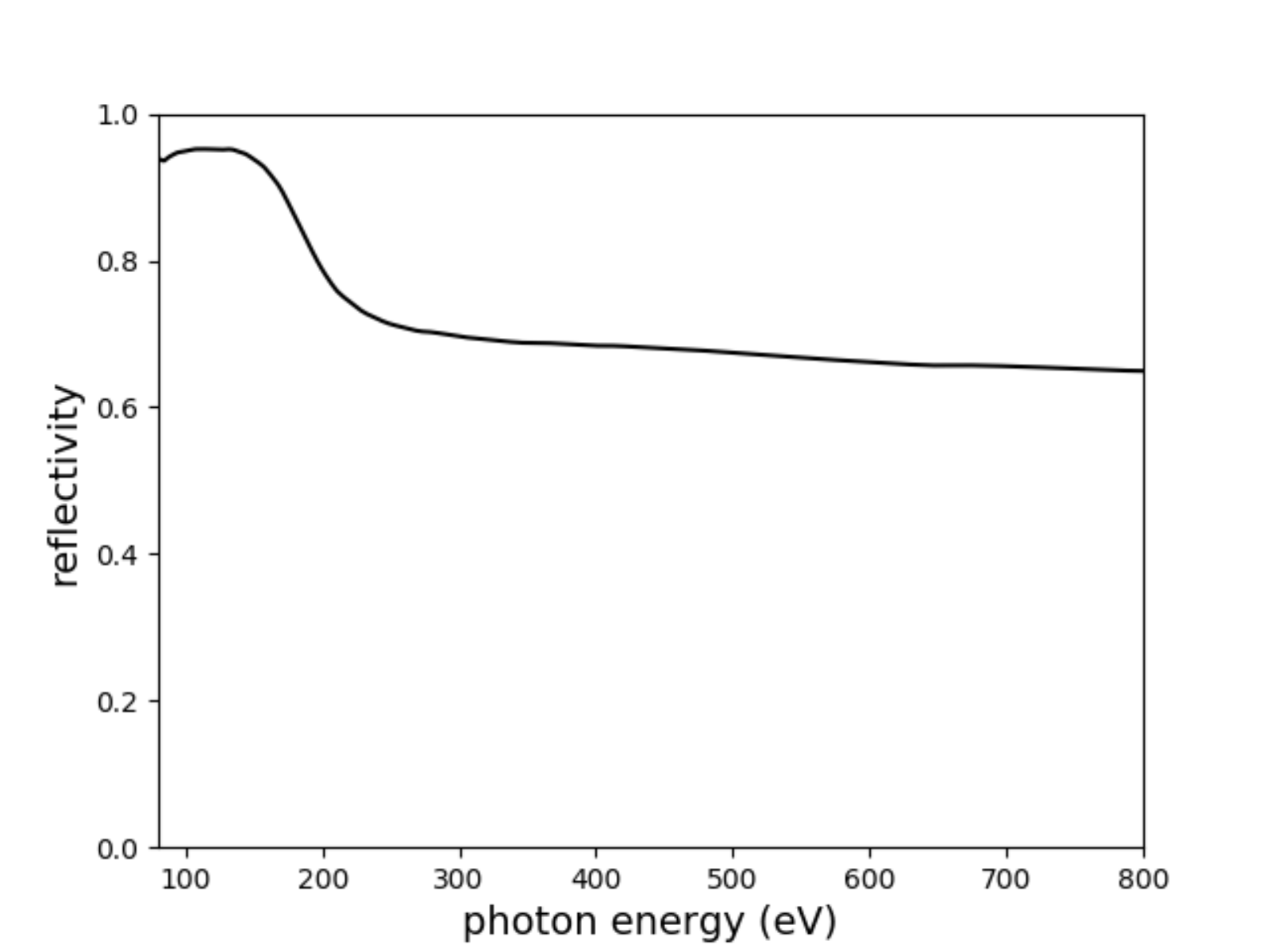}
 \includegraphics[scale=0.55]{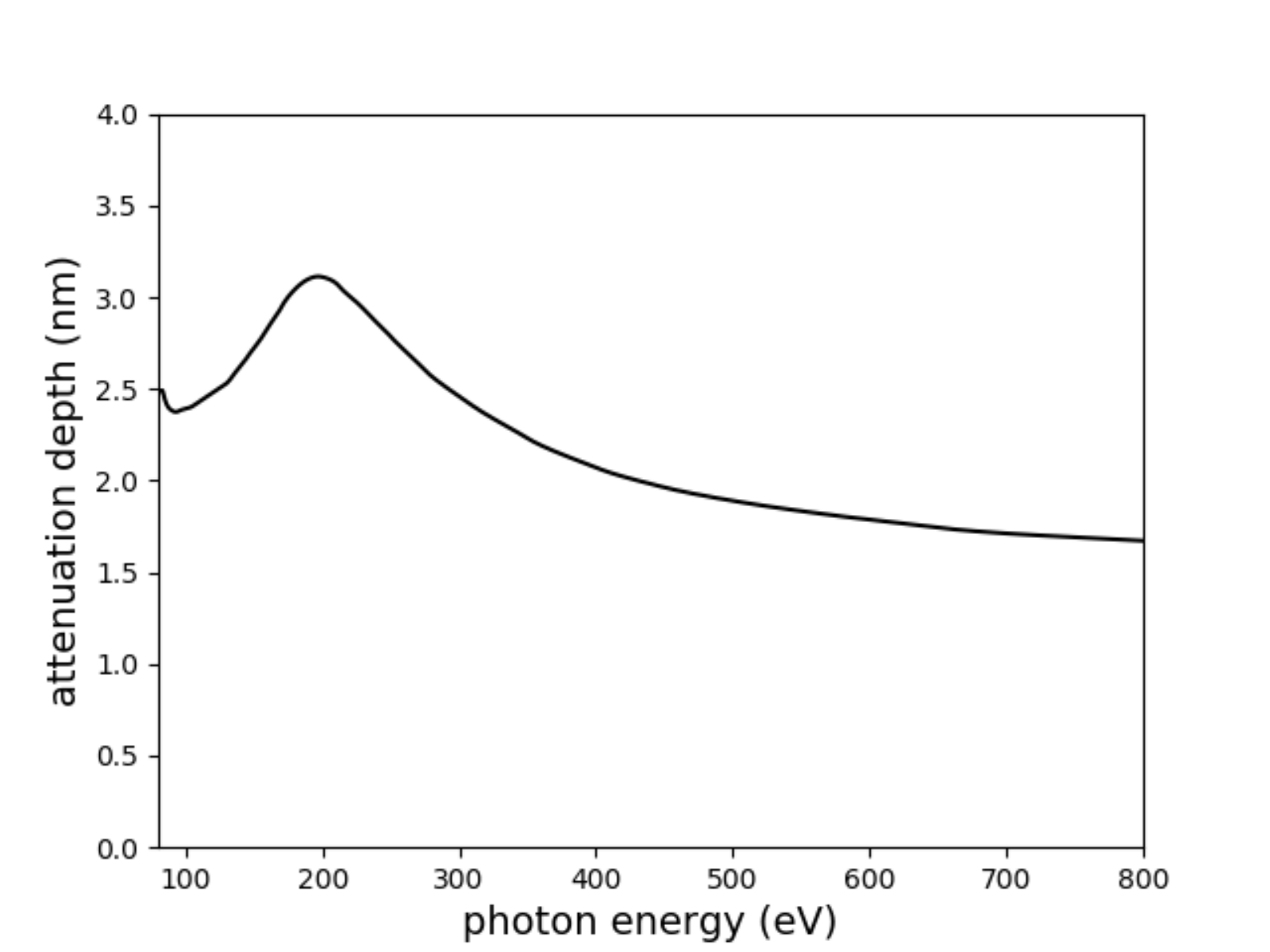}
 \caption{\emph{Left}: Fresnel reflectivity ($\mathcal{R}_F$ given by equation~\ref{eq:fresnel}) for perfectly smooth, thick gold mirror at a $1.7^{\circ}$ grazing incidence angle. \emph{Right}: Attenuation depth ($\mathcal{D}_{\perp}$ given by equation~\ref{eq:attn_depth}) in gold at a $1.7^{\circ}$ grazing incidence angle. Data obtained from the Center for X-ray Optics at Lawrence-Berkeley National Laboratory.}\label{fig:Au_refl_attn}
 \end{figure}

While achieving an effective blaze response from the grating prototype hinges on the shape of the sawtooth facets produced by TASTE, absolute diffraction efficiency is also dependent on the reflectivity of the sawtooth facets at a nominal incidence angle $\zeta \approx 1.7^{\circ}$ as discussed in Section~\ref{sec:als_testing}. 
Having an overcoating on the grating surface relief mold described in Section~\ref{sec:taste_procedure} is important not only for avoiding prominent absorption edges of carbon and oxygen in PMMA but also for preventing potential resist modification by the EUV/SXR beam during diffraction efficiency testing. 
Using data provided by the Center for X-ray Optics (CXRO) at Lawrence Berkeley National Laboratory,\footnote{\url{http://henke.lbl.gov/optical_constants/}} the Fresnel reflectivity of gold at a grazing-incidence angle of $\zeta = 1.7^{\circ}$ is plotted in the left panel of Figure~\ref{fig:Au_refl_attn} as a function of photon energy ranging from 80~eV to 800~eV. 
With $\tilde{\nu}$ as the complex index of refraction of gold, Fresnel reflectivity for a wavefront with transverse electric polarization is given by
\begin{equation}\label{eq:fresnel}
 \mathcal{R}_F = \left\| \frac{\sin \left( \zeta \right) - \sqrt{\tilde{\nu}^2 - \cos^2 \left( \zeta \right)}}{\sin \left( \zeta \right) + \sqrt{\tilde{\nu}^2 - \cos^2 \left( \zeta \right)}} \right\|^2 ,
 \end{equation}
which is approximately equal to the corresponding Fresnel reflectivity for transverse magnetic polarization for grazing-incidence EUV/SXR radiation \citep{Attwood17}. 
Due to this broadband response over the wavelength range for diffraction efficiency testing, 15.5~nm~$> \lambda >$~1.55~nm, gold was chosen as the reflective overcoat for the grating. 
To prevent further reflections at underlaying material interfaces from occurring, the thickness of the gold coating should be chosen appropriately. 
The distance normal to a surface at which radiation loses $1/\mathrm{e}$ of its original intensity is given by the attenuation depth \citep{Gibaud2009} 
\begin{equation}\label{eq:attn_depth}
 \mathcal{D}_{\perp} = \frac{1}{2 \, \text{Im} \! \left[ \tilde{k}_{\perp} \right]} = \frac{\lambda}{4 \pi \, \text{Im} \! \left[ \sqrt{\tilde{\nu}^2 - \cos^2 \left( \zeta \right)} \right]} ,
 \end{equation}
where $\tilde{k}_{\perp} = k_0 \sqrt{\tilde{\nu}^2 - \cos^2 \left( \zeta \right)}$ is the component normal to the surface of the wave vector in gold at a grazing-incidence angle $\zeta$ and $\text{Im} \! \left[ \tilde{k}_{\perp} \right]$ is the imaginary component of $\tilde{k}_{\perp}$. 
From this quantity, which is plotted in the right panel of Figure~\ref{fig:Au_refl_attn} as a function of photon energy using CXRO data, a 15~nm-thick layer deposited on the grating surface relief mold in principle is sufficient to prevent further reflections at underlaying material interfaces from occurring. 
However, because gold is non-reactive toward PMMA, a thin film of an oxidizing metal such as chromium or titanium must be first deposited on the patterned resist to promote wetting and adhesion for the top, reflective layer \citep{Trolier-McKinstry17}. 
Ideally, the result is a gold coating that maintains the fidelity of the sawtooth topography while also realizing blazed groove facets with surface roughness, $\sigma$, low enough to reduce non-specular scatter at EUV and SXR wavelengths as much as possible. 
According to the Fraunhofer criterion for a smooth surface, given by \citep{Beckmann63}
\begin{equation}\label{eq:roughness_inequality}
	\sigma < \frac{\lambda}{32 \sin \left( \zeta \right)} , 
 \end{equation}
$\sigma$ should be on the level of 1~nm RMS to satisfy this condition for 15.5~nm~$> \lambda >$~1.55~nm. 

Deposition for the grating overcoating was performed by electron-beam physical vapor deposition (EBPVD) using a \textsc{Kurt J.\ Lesker Lab-18} system at the PSU Nanofabrication Laboratory. 
First, a 5~nm-thick film of titanium was deposited on the patterned TASTE wafer as described in Section~\ref{sec:taste_procedure} at a previously-determined rate of 0.5~\AA~s$^{-1}$ under high vacuum. 
This allows titanium and some of the oxygen present in PMMA to form a thin oxide layer between the resist surface and the titanium coating, providing a wetted, metallic surface for the gold layer to adhere to. 
Without breaking vacuum, the gold was then deposited at a rate of 1.0~\AA~s$^{-1}$ to achieve a layer $\sim$15~nm thick. 
\begin{figure}
 \centering
 \includegraphics[scale=0.7]{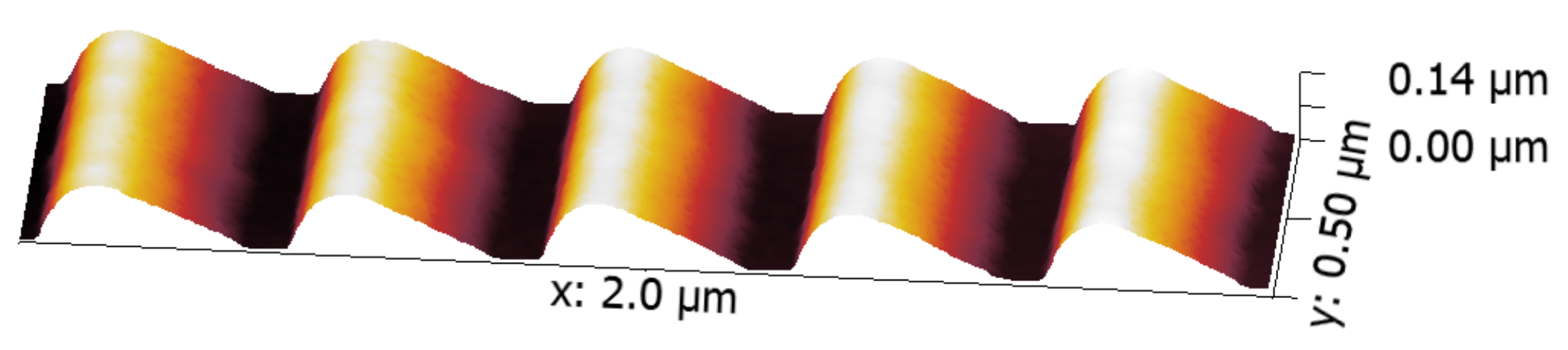}
 \caption{Atomic force micrograph of the grating prototype grooves following electron-beam physical vapor deposition of gold using titanium as an adhesion layer on PMMA.}\label{fig:coating_AFM}
 \end{figure}
The final, coated grating prototype appears under AFM as a sawtooth topography very similar to the uncoated, TASTE-processed resist from Figure~\ref{fig:TASTE_AFMs}. 
This image of the coated grating grooves, taken using the same AFM methodology described in Section~\ref{sec:taste_procedure}, is shown in Figure~\ref{fig:coating_AFM}. 
\begin{figure}
 \centering
 \includegraphics[scale=0.5]{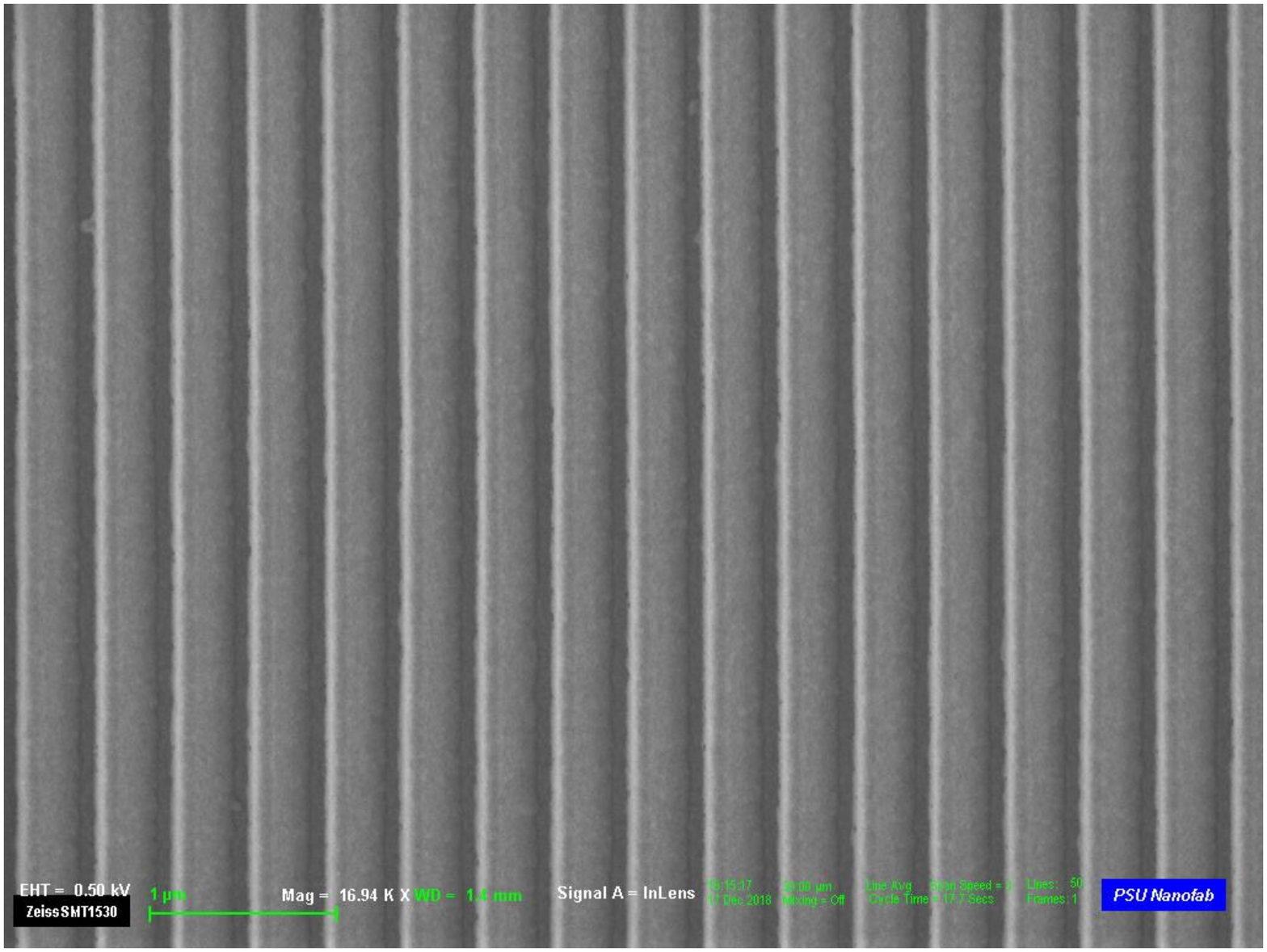}
 \caption{Field emission scanning electron micrograph of the gold-coated grating prototype grooves.}\label{fig:coating_SEM}
 \end{figure}
Moreover, these coated grooves were imaged over a larger area by field emission scanning electron microscopy using a \textsc{Zeiss Leo 1530} system at the Nanofabrication Laboratory of the PSU Materials Research Institute. 
This micrograph, taken at a 0.5~kV electron accelerating voltage, is shown in Figure~\ref{fig:coating_SEM}. 
From the gathered AFM data, $\sigma$ on the groove facets measures about 1.5~nm RMS using the \textsc{NanoScope Analysis} software package provided by \textsc{Bruker} whereas prior to the coating but after thermal reflow, $\sigma \approx 1.25$~nm RMS on PMMA. 
While the blaze angle measures $\delta \approx 27^{\circ}$ as expected, the groove depth measures about 10~nm less than the uncoated resist shown in Figure~\ref{fig:TASTE_AFMs}. 
Moreover, the bottom plateau of the coated grooves appears slightly widened relative to the bottom plateau of the bare, TASTE-processed resist, where the surface of the silicon substrate is exposed, suggesting that the EBPVD process produces a thicker metal coating on a silicon surface with native oxide than it does on PMMA resist. 
However, because these regions are to a high degree shadowed to the incoming radiation in a near-Littrow configuration, this is not expected to have a large impact on diffraction efficiency. 

\section{Testing Results}\label{sec:results}
\begin{figure}
 \centering
 \includegraphics[scale=0.65]{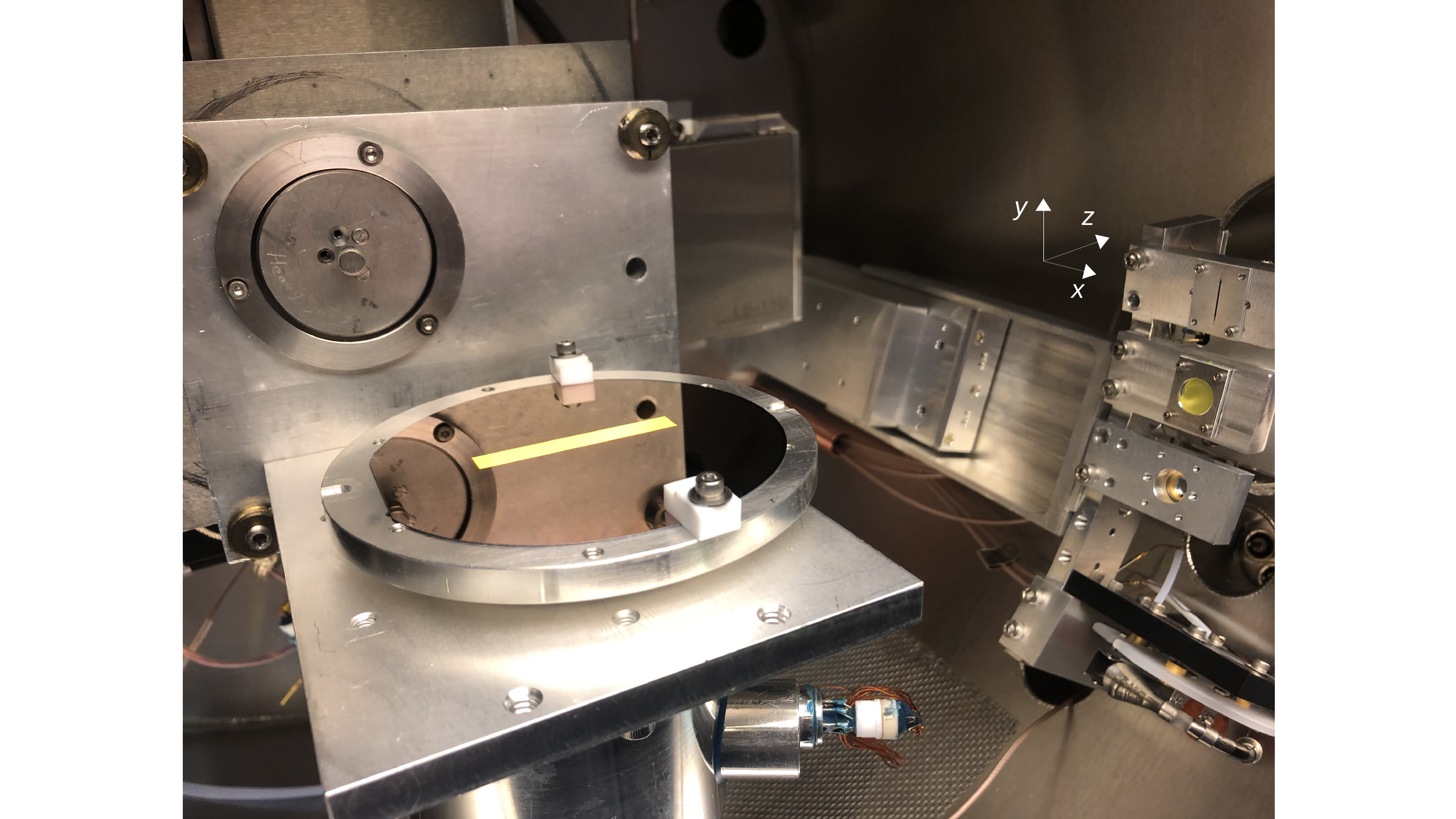}
 \caption{Grating prototype installed inside the test chamber of beamline 6.3.2 for extreme ultraviolet and soft x-ray reflectometry at the Advanced Light Source of Lawrence-Berkeley National Laboratory.}\label{fig:test_chamber} 
 \end{figure}

Following the methodology outlined in Section~\ref{sec:als_testing} and detailed by \citet{Miles18}, the grating prototype was tested for EUV and SXR diffraction efficiency at beamline 6.3.2 of the ALS. 
Figure~\ref{fig:test_chamber} shows the gold-coated grating prototype installed inside the beamline test chamber in an extreme off-plane mount, where the dispersion direction, $x$, is roughly parallel with the direction of horizontal stage motion for the photodiode detector, which is seen masked with a 0.5~mm-wide vertical slit. 
The grating was first oriented at a yaw angle of $\varphi \approx 0^{\circ}$ with graze and roll angles, $\eta$ and $\phi$ respectively, being approximately zero as measured by the tilt of the optic mount using a spirit level. 
The graze angle was then adjusted to the nominal test value of $1.5^{\circ}$ by using the goniometric stage motion of the photodiode to ensure that the angle between the direct beam and the reflected beam is roughly $2 \eta \approx 3^{\circ}$. 
Next, all grating geometric angles introduced in Section~\ref{sec:als_testing} were determined experimentally through analyzing the arc of diffraction as sampled by the photodiode. 
While the $x$-positions of propagating orders can be determined by sampling the diffracted arc along the horizontal direction of the detector staging, their positions along the cross-dispersion direction, $y$, require knowledge of the system throw, $L \approx 235$~mm, to map the goniometric angle associated with the stage motion, $\Theta$, to a $y$-coordinate using $y = L \sin \left( \Theta \right)$. 
Thus with a measured value for $L$, the diffracted arc can be fit to a circle to determine values for the arc radius, $r$, as well as the $x-y$ coordinates of the arc center. 
By comparing these to the positions of the direct beam and $0^{\text{th}}$ order as they fall on the diffracted arc, the orientation of the grating relative to the incident beam and the photodiode staging could be determined experimentally. 
From these measurements, the grating was set to a near Littrow configuration by adjusting $\varphi$ to ensure that $\alpha \approx 27^{\circ}$ and $\gamma \approx 1.7^{\circ}$ at a graze angle of $\eta \approx 1.5^{\circ}$. 

The throw of the system at the location of $0^{\text{th}}$ order was measured to be $L = 233.0 \pm 1.4$~mm by comparing the known detector length of 10~mm to the angular size of the detector as measured by a goniometric scan of the beam. 
In principle, $L$ changes as the the detector moves along the direction $x$ with focal corrections on the order of tens of micrometers within 10~mm of travel. 
However, for the analysis and discussion that follows, these corrections are ignored so that order locations are mapped using $x$ and $y = L \sin \left( \Theta \right)$  with $L$ fixed at the measured value. 
In the final test geometry, the diffracted arc was mapped using data gathered at 450~eV and 500~eV in steps of 50~$\micron$ along the $x$-direction of the photodiode staging. 
\begin{figure}
 \centering
 \includegraphics[scale=0.45]{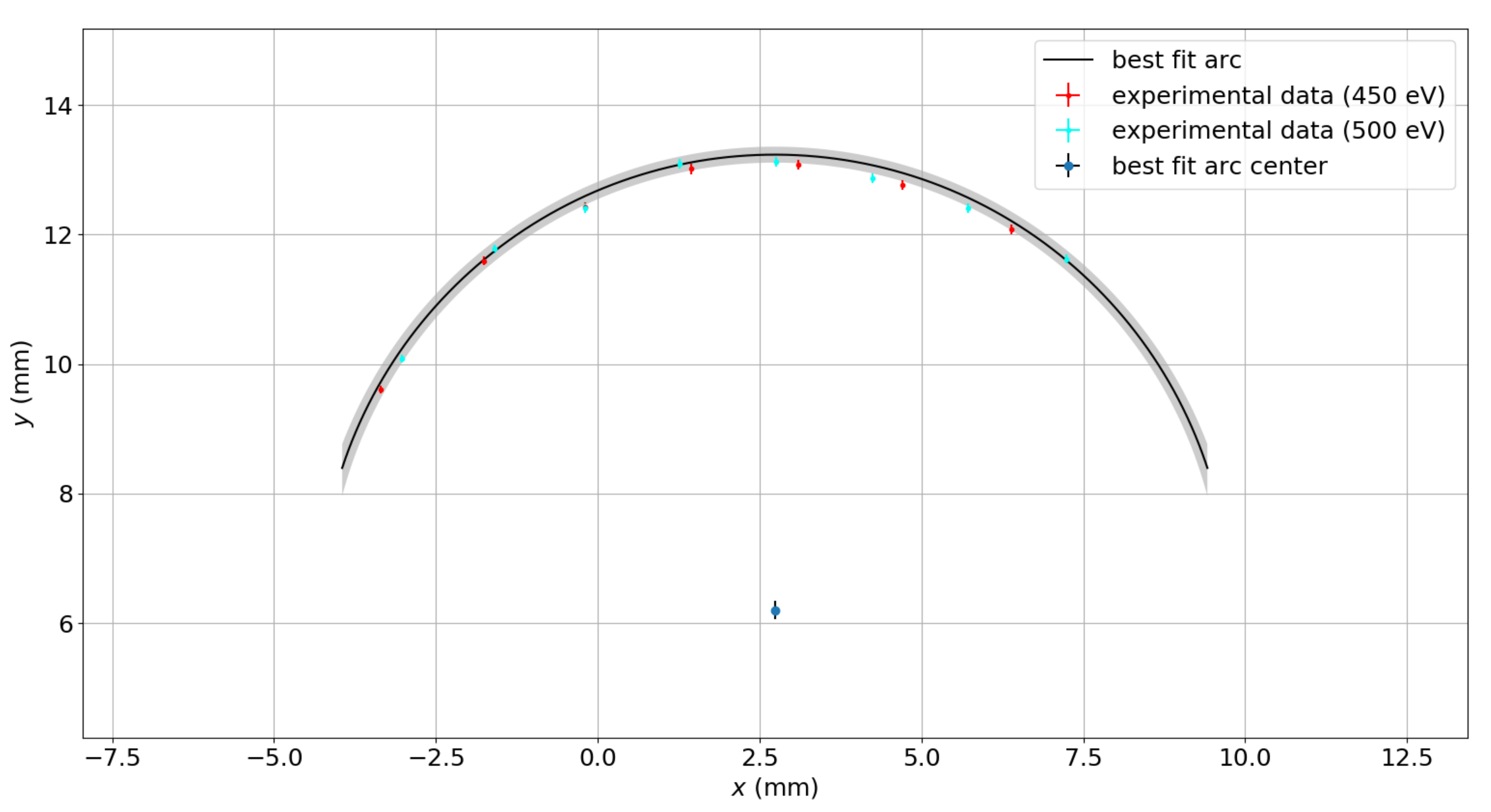}
 \caption{Diffracted arc for the beamline test configuration mapped using data gathered at 450~eV and 500~eV and fit to a circle. Grayed regions represent one standard deviation uncertainty.}\label{fig:arc_fit}
 \end{figure}
By fitting these data to a half-circle as shown in Figure~\ref{fig:arc_fit}, the arc radius was measured as $r = 7.03 \pm 0.12$~mm and from $r = L \sin \left( \gamma \right)$ by equation~\ref{eq:arc_radius}, the cone opening half-angle for the diffraction pattern was determined to be $\gamma = 1.73 \pm 0.03^{\circ}$. 
Next, the azimuthal incidence angle, $\alpha$, was measured independently of the roll angle using 
\begin{equation}\label{eq:exp_alpha}
 \sin \left( \alpha \right) = \frac{\Delta x_{\text{dir}}}{r} ,
 \end{equation}
where $\Delta x_{\text{dir}}$ is the $x$-distance between the direct beam (not shown in Figure~\ref{fig:arc_fit}) and the center of the diffracted arc determined from the fit. 
With a measured value of $\alpha = 23.4 \pm 0.6^{\circ}$, the roll angle was constrained as $\phi = 1.14 \pm 0.04^{\circ}$ using
\begin{equation}\label{eq:exp_roll}
 \sin \left( \phi \right) = \frac{\Delta x_0}{r} - \sin \left( \alpha \right) , 
 \end{equation}
where $\Delta x_0$ is the $x$-distance between $0^{\text{th}}$ order and the center of the diffracted arc. 
Using this result and $\Delta y_0$, the $y$-distance between $0^{\text{th}}$ order and the center of the diffracted arc, a graze angle of $\eta \approx 1.5^{\circ}$ was verified through 
\begin{equation}\label{eq:exp_eta}
 \sin \left( \eta \right) = \frac{\Delta y_0}{L \cos \left( \phi \right)}
 \end{equation}
to give $\eta = 1.56 \pm 0.04^{\circ}$. 
\begin{table}[]
\centering
\begin{tabular}{@{}ll@{}}
\toprule
parameter                                                                 & measured value             \\ \midrule
system throw ($L$)                                                        & $232.0 \pm 1.4$ mm     \\
arc radius ($r$)                                                          & $7.03 \pm 0.12$ mm       \\
$x$-distance between direct beam and arc center ($\Delta x_{\text{dir}}$) & $2.80 \pm 0.05$ mm       \\
$x$-distance between 0$^{\text{th}}$ order and arc center ($\Delta x_0$)  & $2.92 \pm 0.05$ mm       \\
$y$-distance between 0$^{\text{th}}$ order and arc center ($\Delta y_0$)  & $6.33 \pm 0.14$ mm       \\ \midrule
cone opening half-angle ($\gamma$) by equation~\ref{eq:arc_radius}        & $1.73 \pm 0.03^{\circ}$  \\
azimuthal incidence angle ($\alpha$) by equation~\ref{eq:exp_alpha}       & $23.4 \pm 0.6^{\circ}$ \\ \midrule
roll (rotation about $z$-axis; $\phi$) by equation~\ref{eq:exp_roll}      & $1.14 \pm 0.04^{\circ}$  \\
graze (rotation about $x$-axis; $\eta$) by equation~\ref{eq:exp_eta}      & $1.56 \pm 0.04^{\circ}$  \\
yaw (rotation about $y$-axis; $\varphi$) by equation~\ref{eq:exp_yaw}     & $0.69 \pm 0.01^{\circ}$  \\ \bottomrule
\end{tabular}
\caption{Measured parameters for the diffracted arc in test configuration at beamline 6.3.2 of the Advanced Light Source.}\label{tab:arc_params}
\end{table}
Finally, grating yaw was measured using 
\begin{equation}\label{eq:exp_yaw}
 \sin \left( \varphi \right) = \frac{\Delta x_{\text{dir}}}{L \cos \left( \eta \right)}
 \end{equation}
to yield $\varphi = 0.69 \pm 0.01^{\circ}$. 
Summarized in Table~\ref{tab:arc_params}, these measurements indicate a near-Littrow test configuration at $\eta \approx 1.5^{\circ}$ for a blaze angle of $\delta \approx 27^{\circ}$.

In the test geometry discussed above, diffraction efficiency data were gathered as a function of photon energy where for each measurement, both the diffracted arc and the direct beam were scanned along the $x$-direction in 50~$\micron$ increments. 
From these measurements, absolute diffraction efficiency (\emph{i.e.}, $\mathcal{E}_n = \mathcal{I}_n / \mathcal{I}_{\text{inc}}$ given by equation~\ref{eq:diffraction_efficiency}) was calculated by identifying the maximum of each diffracted order and the direct beam, taking three intensity measurements around the centroid and then dividing each order by the direct beam after subtracting out the appropriate noise floors. 
Contributions to this noise include dark current from the photodiode detector, which was measured using the photodiode readout in the absence of the EUV/SXR beam, and additionally, diffuse scatter arising from surface roughness on the groove facets. 
The latter, which in principle only affects $\mathcal{I}_n$, was estimated using the continuum level in between order maxima and was found to be significant only for photon energies starting at 600~eV where it contributed to $\mathcal{E}_n$ on the level of a percent or less for each propagating order. 
These diffraction efficiency measurements were performed in 20~eV steps, first from 440 to 800 eV and then from 80 to 420 eV. 
For the latter set of measurements, the triple-mirror order sorter at the beamline is required to maintain a spectrally pure beam provided by the monochromator \citep{Gullikson01}.  
\begin{figure}
 \centering
 \includegraphics[scale=0.45]{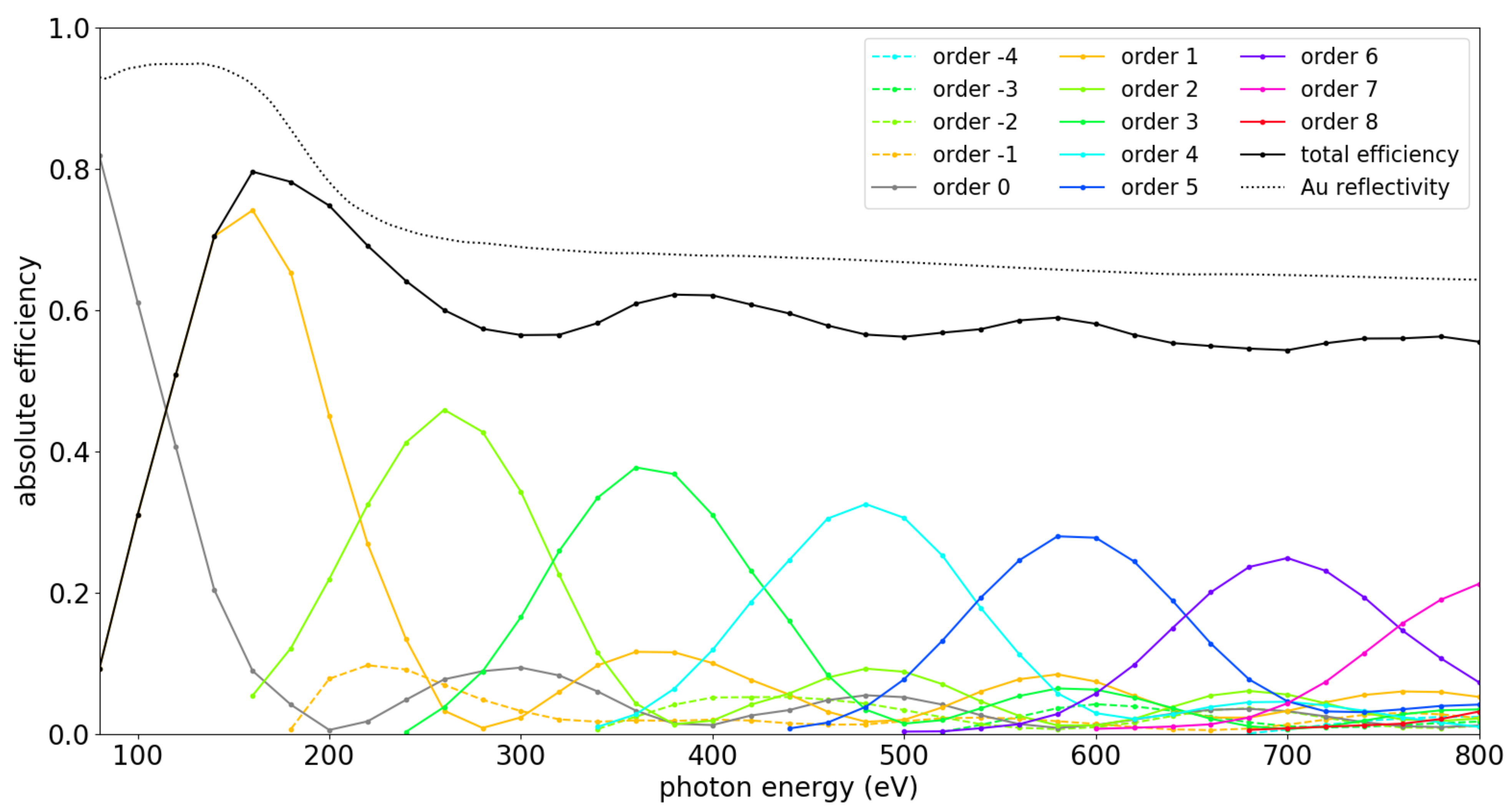}
 \caption{Absolute diffraction efficiency measurements taken at the Advanced Light Source compared to the Fresnel reflectivity of gold, $\mathcal{R}_{F}$. Total diffraction efficiency below 240~eV misses contributions from orders 2 and 3 on the order of a few percent.}\label{fig:abs_eff}
 \end{figure} 
Although the implementation of the order sorter is expected to shift slightly the position of the beam on the grating, and hence the measured parameters listed in Table~\ref{tab:arc_params}, the effect is small and not apparent in the measured absolute efficiency data, which are shown in Figure~\ref{fig:abs_eff} compared to the Fresnel reflectivity of gold, $\mathcal{R}_{F}$, at $\zeta = 1.73^{\circ} \approx \gamma$ using equation~\ref{eq:fresnel}. 
However, as indicated most clearly by the sharp cut-off in the measured $n=2$ curve at 160~eV, the beam shift evidently caused measurements of propagating orders of $n=2$ and $n=3$ with large diffracted angle, $\beta$, to be missed by the photodiode during data collection. 
These data nonetheless show that peak order efficiency ranges from about 75\% down to 25\% as photon energy, and order number, increase. 
The total diffraction efficiency, defined as $\mathcal{E}_{\text{tot}} \equiv \sum_n \mathcal{E}_n$ for all propagating orders with $n \neq 0$, is also plotted in Figure~\ref{fig:abs_eff} but due to the missing $n=2$ and $n=3$ measurements in the EUV, this curve underestimates the true total diffraction efficiency for photon energies smaller than 240~eV. 
Moreover, relative efficiency was calculated by dividing each $\mathcal{E}_n$ measurement from Figure~\ref{fig:abs_eff} by $\mathcal{R}_{F}$. 
\begin{figure}
 \centering
 \includegraphics[scale=0.45]{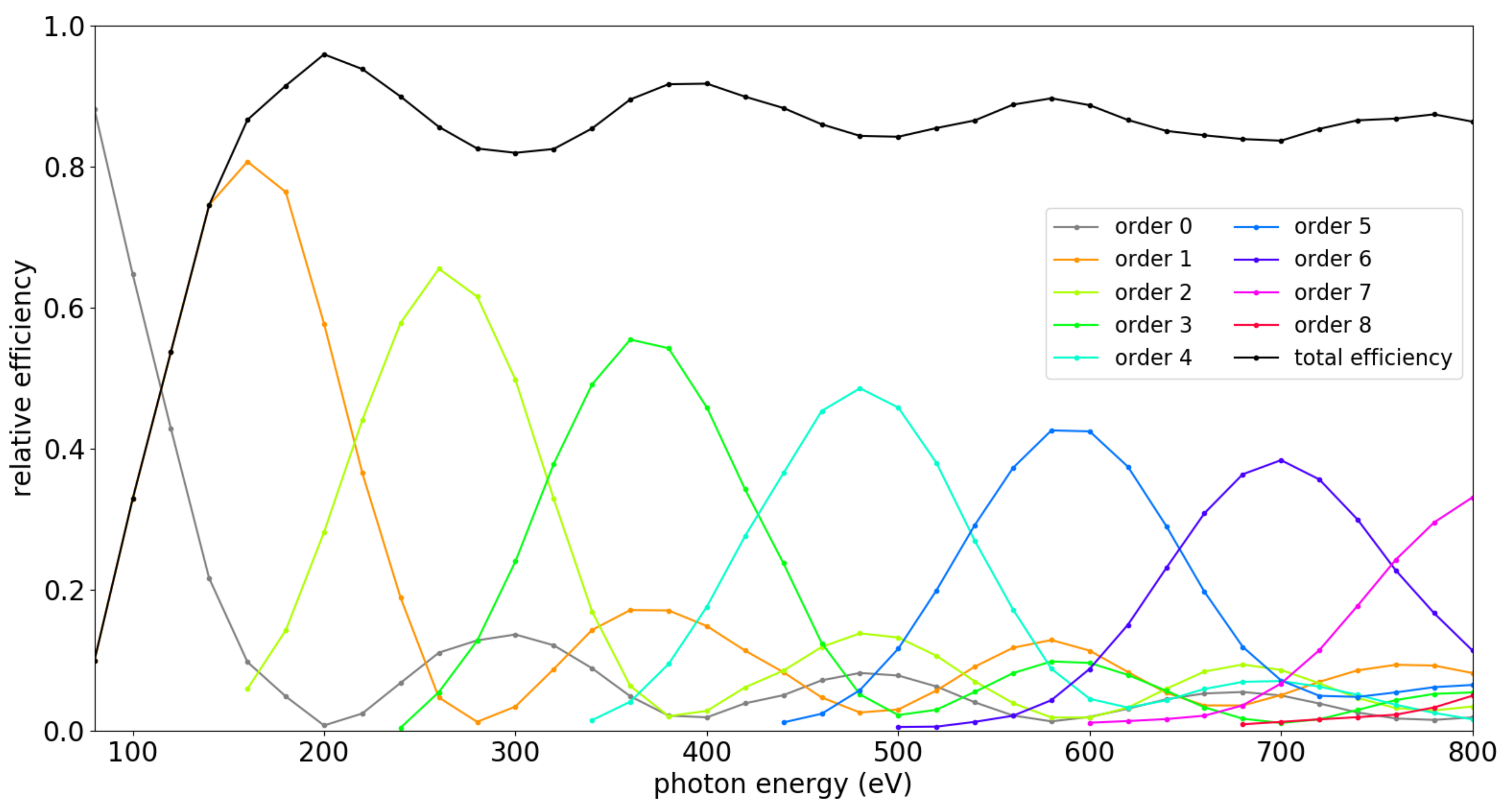}
 \caption{Relative diffraction efficiency calculated by dividing the absolute diffraction efficiency from Figure~\ref{fig:abs_eff} by the Fresnel reflectivity of gold, $\mathcal{R}_{F}$. Total diffraction efficiency below 240~eV misses contributions from orders 2 and 3 on the order of a few percent.}\label{fig:rel_eff}
 \end{figure}
This result is plotted in Figure~\ref{fig:rel_eff} where total relative diffraction efficiency, $\mathcal{E}_{\text{tot}} / \mathcal{R}_{F}$, ranges from about 95\% to 88\% as photon energy increases from 240~eV to 800~eV, where all propagating orders are accounted for. 

\section{Discussion}\label{sec:discussion}
The beamline measurements presented in Section~\ref{sec:results} indicate that the grating prototype yields an approximate blaze response at EUV and SXR wavelengths in a near-Littrow configuration.  
This is evidenced by the total diffraction efficiency shown in Figures \ref{fig:abs_eff} and \ref{fig:rel_eff} being dominated by single orders with positive $n$ and peak positions close to those predicted by equation~\ref{eq:blaze_wavelength_littrow} for the blaze wavelength. 
However, along with the peak orders that resemble a blaze response, propagating orders of lower $n$ each contribute to the total absolute efficiency at a level of about 10\%. 
Thus toward the blue end of the measured bandpass where a relatively large number of propagating orders exist by equation~\ref{eq:off-plane_incidence_orders}, peak-order diffraction efficiency is comparatively low and comprises a smaller fraction of $\mathcal{E}_{\text{tot}}$. 
This suggests that diffracted orders gradually become suppressed with increasing $n$ due to an imperfect sawtooth topography generated by the TASTE process outlined in Section~\ref{sec:taste_procedure}. 
That is, while an idealistic blazed grating exhibits a sharp sawtooth topography, the grating prototype features a quasi-flat apex produced by the 100~nm-wide, top staircase step in the GEBL pattern that is nominally unexposed to high-energy electrons and hence largely unaffected by the thermal reflow process. 

\begin{figure}
 \centering
 \includegraphics[scale=0.45]{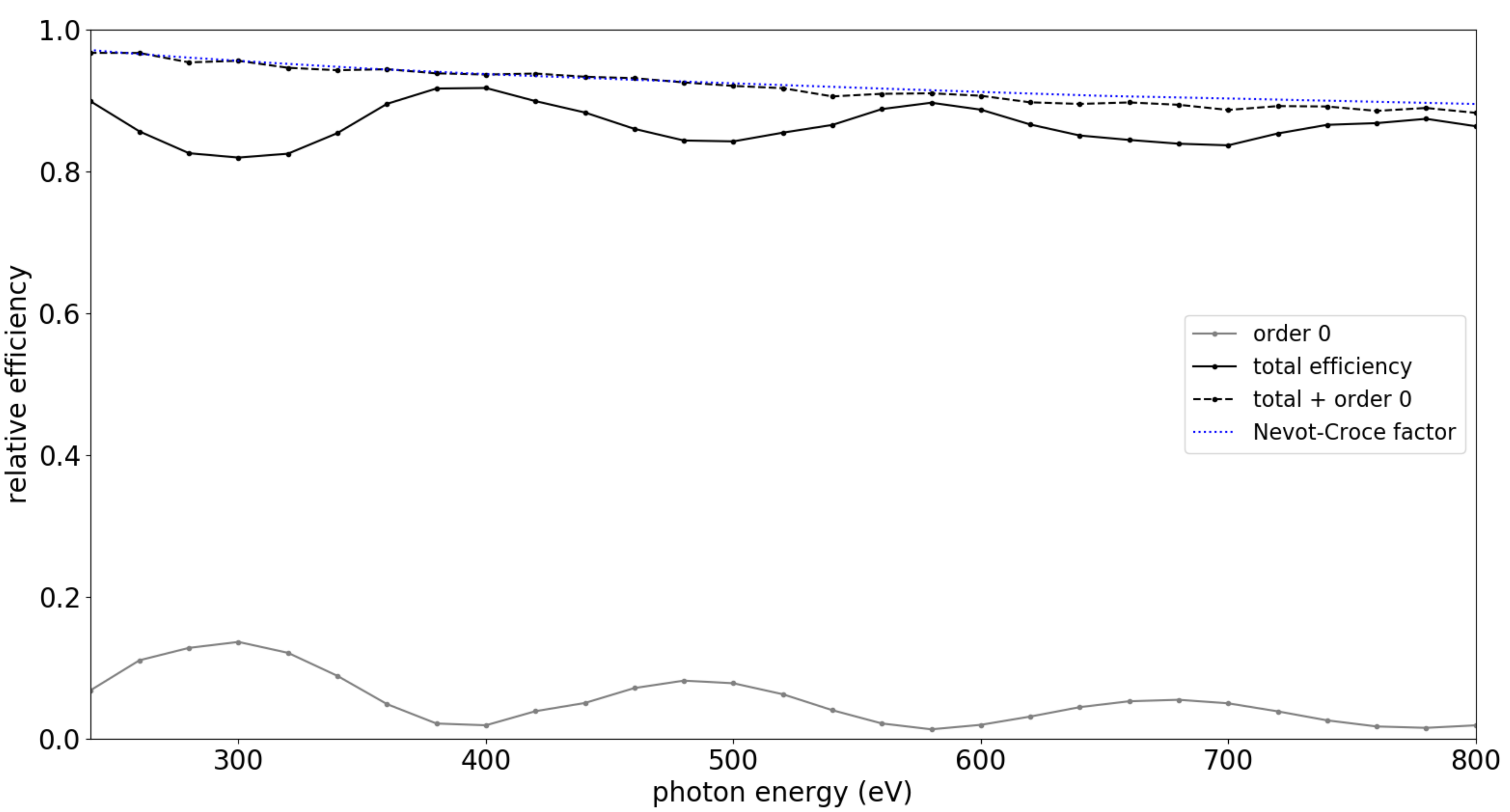}
 \caption{Total grating response, defined as the sum of total diffraction efficiency and zero order, plotted relative to the reflectivity of gold. Overlaid is the Nevot-Croce factor given by equation~\ref{eq:nc_factor} for $\zeta = 1.73^{\circ}$ and $\sigma = 1.5$~nm RMS, which indicates the theoretical specular reflectivity of a rough surface relative to Fresnel reflectivity.}\label{fig:rough_eff}
 \end{figure}

In addition to an imperfect sawtooth topography, peak-order diffraction efficiency, especially toward the blue end of the spectrum, is impacted by $\lambda$-dependent losses that arise from surface roughness on the groove facets. 
This is gleaned from analyzing the total relative response from the grating, defined as $\left( \mathcal{E}_{\text{tot}} + \mathcal{E}_0 \right) / \mathcal{R}_{F}$, where $\mathcal{R}_{F}$ is the Fresnel reflectivity of gold at the angle $\zeta$ introduced in Section~\ref{sec:als_testing}. 
Due to the short, nano-scale attenuation depth of gold at grazing incidence as discussed in Section~\ref{sec:coating}, it is justified to treat the grating overcoat material as an infinitely-thick layer of gold using EUV and SXR optical constants provided by CXRO.
The grating's total relative response is plotted in Figure~\ref{fig:rough_eff} over the range of measured photon energies that include all propagating orders, where the data show a monotonic decrease from about 96\% down to 88\% as wavelength decreases, suggesting that $\lambda$-dependent losses are occurring. 
This is to be compared with the specular reflectivity of a hypothetical mirror flat relative to $\mathcal{R}_{F}$ such that its total relative response is 100\% in the absence of surface roughness. 
In the regime of total external reflection, the reduced specular reflectivity from a rough surface, $\mathcal{R}_{\text{rough}}$, is described approximately by the Nevot-Croce factor \citep{Nevot80,deBoer95,Gibaud2009}. 
For a thick slab of gold with complex index of refraction $\tilde{\nu}$ and RMS surface roughness $\sigma$, this factor is given by 
\begin{equation}\label{eq:nc_factor}
 \frac{\mathcal{R}_{\text{rough}}}{\mathcal{R}_{F}} = \left\| \mathrm{e}^{-2 k_{\perp} \tilde{k}_{\perp} \sigma^2 } \right\|^2 = \mathrm{e}^{-4 \sigma^2 k_0^2 \sin \left( \zeta \right) \, \text{Re} \left[ \sqrt{\tilde{\nu}^2 - \cos^2 \left( \zeta \right)} \right]} ,
 \end{equation}
where, as described in Section~\ref{sec:coating}, $k_{\perp} = k_0 \sin \left( \zeta \right)$ and $\tilde{k}_{\perp} = k_0 \sqrt{\tilde{\nu}^2 - \cos^2 \left( \zeta \right)}$ are the components of the wave vector normal to the surface in vacuum and gold, respectively, with $k_0 \equiv 2 \pi / \lambda$. 
Moreover, $\text{Re} \! \left[ \sqrt{\tilde{\nu}^2 - \cos^2 \left( \zeta \right)} \right]$ represents the real part of $\tilde{k}_{\perp} / k_0$. 

As described by \citet{deBoer95}, the Nevot-Croce factor defined by equation~\ref{eq:nc_factor} is valid for small roughness features taking on a Gaussian height distribution with $k_{\perp} \sigma \ll 1$ so that using $\sigma \approx 1.5$~nm RMS as measured by AFM and $\zeta = 1.73^{\circ}$, this condition is satisfied for  
\begin{equation}\label{eq:rough_condition}
 \lambda \gg 2 \pi \sigma \sin \left( \zeta \right) \approx 0.3 \text{ nm} .
 \end{equation}
Additionally, derivations of the Nevot-Croce factor assume a surface correlation length, $\xi$, satisfying $\xi k^2_{\perp} \ll k_0$. 
Keeping $\xi$, which represents the lateral size scale of roughness features, as an unknown, this yields 
\begin{equation}\label{eq:corr_condition}
 \lambda \gg 2 \pi \xi \sin^2 \left( \zeta \right) \approx 0.006 \xi . 
 \end{equation}
If equations \ref{eq:rough_condition} and \ref{eq:corr_condition} are fulfilled, diffuse scatter in vacuum can in principle be neglected and $\lambda$-dependent losses attributed to absorption as radiation scatters into the medium. 
Otherwise, radiation of wavelength $\lambda$ is able to diffract from roughness spatial frequencies on the order of $\xi^{-1}$, producing diffuse scatter that can be detected by the photodiode, in which case details of the roughness power spectrum are required to obtain a more accurate expression for $\mathcal{R}_{\text{rough}}$ \citep{deBoer95,Wen15}. 
Because this information is not known for the groove facets on the grating prototype, equation~\ref{eq:nc_factor} was taken to approximate the total relative response from the grating prototype in the presence of surface roughness. 
This is plotted in Figure~\ref{fig:rough_eff}, where it is seen that the data closely match the Nevot-Croce factor with the experimentally-determined values of $\zeta \approx \gamma = 1.73^{\circ}$ and $\sigma \approx 1.5$~nm RMS. 
This supports the idea that surface roughness on the groove facets is responsible for the losses in the grating's total response over the measured bandpass that encompasses all propagating orders. 
Although the detection of diffuse scatter for photon energies 600~eV and higher as described in Section~\ref{sec:results} suggests that the conditions for the Nevot-Croce factor to be valid are not strictly fulfilled at these relatively short wavelengths, Figure~\ref{fig:rough_eff} indicates that equation~\ref{eq:nc_factor} is a decent approximation across the bandpass considered.
However, future diffraction efficiency test campaigns should better quantify diffuse scatter due to surface roughness in a similar manner to x-ray reflectivity experiments that aim to characterize surfaces, materials and interfacial roughness \citep{Gay1999,Baumbach1999}. 

To investigate the impact that an imperfect sawtooth topography with an unpointed apex has on the measured diffraction efficiency, absolute diffraction efficiency was modeled according to vector diffraction theory. 
This was handled using the software package \textsc{PCGrate-SX} version 6.1,\footnote{\url{https://www.pcgrate.com/loadpurc/download}} which solves the Helmholtz equation through the integral method for a custom grating boundary and incidence angles input by the user \citep{Goray10}. 
Based on the findings of \citet{Marlowe16}, which verify a lack of polarization sensitivity for SXR gratings used in extreme off-plane mounts, \textsc{PCGrate-SX} calculations were carried out assuming a perfectly conducting grating boundary with perfectly smooth groove facets and an incident wavefront with transverse electric polarization.  
While perfect conductivity combined with the absence of surface roughness implies a lossless response from the grating grooves, \textsc{PCGrate-SX} modulates the predicted diffraction efficiency by the reflectivity of a user-input, stratified medium defined by optical constants and custom layer thicknesses. 
Taking the grating material to be an infinitely-thick layer of gold as discussed above, the edge-on groove shape of the grating prototype was approximated as an acute trapezoid with a near-vertical lateral side opposite a slope that emulates the active blaze facet. 
Additionally, a flat bottom portion was included to represent the cleared portion of the resist described in Section~\ref{sec:taste_procedure}. 
\begin{figure}
 \centering
 \includegraphics[scale=0.75]{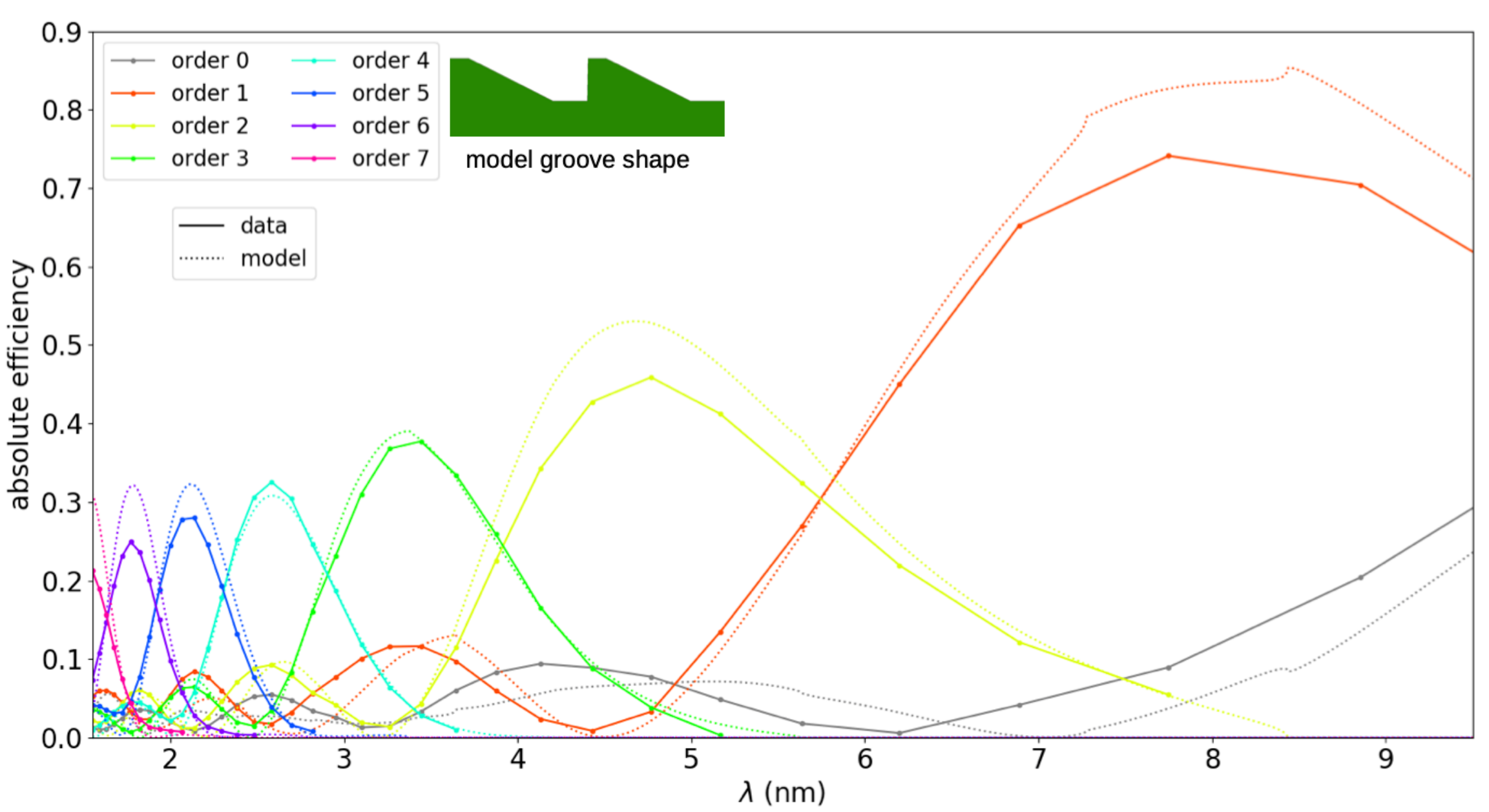} 
 \caption{Absolute diffraction efficiency from Figure~\ref{fig:abs_eff} compared to theoretical diffraction efficiency modeled using the \textsc{PCGrate-SX} software package. Modeled data are multiplied by the Nevot-Croce factor given by equation~\ref{eq:nc_factor} for $\zeta = 1.73^{\circ}$ and $\sigma = 1.5$~nm RMS.}\label{fig:model_eff}
 \end{figure}
Using the nominal values of $\alpha$ and $\gamma$ listed in Table~\ref{tab:arc_params} for grating incidence angles and $d=400$~nm for the groove spacing, a series of \textsc{PCGrate-SX} calculations were performed for a range of trapezoids with slightly-varying dimensions close to those measured by AFM in Figure~\ref{fig:coating_AFM}. 
The model matching the measured data most closely was one with a blaze angle of $\delta = 27^{\circ}$, a groove depth of 120~nm, a flat-top width of 77~nm and bottom-width of 85~nm. 
These predicted data, modulated by the Nevot-Croce factor from Figure~\ref{fig:rough_eff}, are plotted as a function of $\lambda$ in Figure~\ref{fig:model_eff} and compared to the measured diffraction efficiency, $\mathcal{E}_n$, for orders $n=0$ through $n=7$. 

Figure~\ref{fig:model_eff} shows that the measured peak order positions match roughly those predicted by the model, demonstrating that the grating prototype has an efficiency response similar to that of a blazed grating with $d=400$~nm and $\delta = 27^{\circ}$ at the experimentally-determined incidence angles of $\alpha = 23.4^{\circ}$ and $\gamma = 1.73^{\circ}$. 
However, the amplitudes of the peak orders generally fall short of the model with the apparent exceptions of $n=3$ and $n=4$. 
This phenomenon seems to be due in part to the mismatches that exist between the measured data and the model for secondary diffraction peaks, suggesting that the groove shape trapezoidal approximation is not sufficient to reproduce these results to a high degree of accuracy. 
The grating prototype grooves likely have an apex that is slightly rounded as a result of the thermal reflow process but this is difficult to verify through AFM because the shape of the \textsc{SCANASYST-AIR} tip is convolved with the true grating topography in the micrographs shown in Figures \ref{fig:TASTE_AFMs} and \ref{fig:coating_AFM}. 
Nonetheless, rounding of corners or other deviations from an ideal acute trapezoid are expected to have an impact on the distribution of diffraction efficiency among orders. 
Figure~\ref{fig:model_eff} also shows that measured peak order efficiency becomes increasingly diminished relative to the model as order number increases beyond $n=4$, which is consistent with the observation already mentioned that the relatively large number of orders at short $\lambda$ each contribute substantially to the total efficiency, $\mathcal{E}_{\text{tot}}$, while the peak order comprises a relatively smaller fraction. 
This is another indication of there being groove shape imperfections that diminish the grating prototype's blaze response. 
A possible explanation beyond rounded corners at the apex is irregularity or non-flatness of the sloped surfaces of the grating grooves across the prototype. 
In principle this could be caused in part by a non-uniform spin coat thickness but it is expected that the imperfect blazed grating topography produced by TASTE process described in Section~\ref{sec:taste_procedure} is the largest contributor to this issue.  

\section{Summary and Conclusions}\label{sec:summary}
A prototype for a reflection grating with a groove spacing of 400~nm was fabricated at the PSU Materials Research Institute by generating an approximate sawtooth topography in 130~nm-thick PMMA resist coated on a silicon wafer through the process of TASTE and then coating the grating grooves with a thin layer of gold via EBPVD for EUV and SXR reflectivity. 
Diffraction efficiency measurements spanning 15.5~nm~$> \lambda >$~1.55~nm in a grazing incidence, extreme off-plane mount collected at beamline 6.3.2 of the ALS demonstrate that the grating behaves approximately as a blazed grating with groove cross-sections shaped like an acute trapezoid and a blaze angle of $\delta \approx 27^{\circ}$. 
The total response from the grating relative to the reflectivity of the gold overcoat measures between 96\% and 88\% in the SXR, with losses attributed to absorption and diffuse scatter from grating facets with 1.5~nm RMS surface roughness. 
However, even with losses accounted for, the blaze response is observed to diminish for peak orders with $n=5$ and greater. 
While this phenomenon is a result of the TASTE process yielding an imperfect sawtooth topography, these results show that TASTE is a promising fabrication technique for the manufacture of custom reflection gratings for EUV and SXR spectroscopy.

An especially important feature of the TASTE process is its ability to define a sawtooth-like topography over a layout defined by electron-beam lithography while also avoiding the dependences on crystallographic structure that exist in processes that use anisotropic wet etching to provide a grating blaze \citep{Franke97,Chang03,McEntaffer13,Miles18}. 
This is particularly advantageous for realizing fanned, curved or other variable-line-space groove layouts that are required for achieving high spectral resolving power, $\lambda / \Delta \lambda$, while also having blazed groove facets that enable high spectral sensitivity. 
With total absolute diffraction efficiency exceeding 40\% in the SXR bandpass, these results show that gratings fabricated by TASTE are capable of meeting \emph{Lynx} requirements in terms of spectral sensitivity. 
Additionally, an absolute efficiency of 75\% in first order at 160~eV gives an indication that TASTE can realize a highly-efficient grating for EUV spectroscopy with modification of grating parameters. 
However, further work in nanofabrication and spectral resolving power testing is required to determine to what degree TASTE is able to make improvements in these areas of technological development. 
In particular, producing gratings with groove spacing significantly smaller than 400~nm that maintain a satisfactory sawtooth topography is challenging from the standpoint of fabrication by TASTE. 
This last item is crucial for SXR reflection gratings that often call for groove spacings near 160~nm and is the subject of a forthcoming publication \citep{McCurdy19}. 

\acknowledgments
This research was supported by NASA Space Technology Research Fellowship under grant no.\ NNX16AP92H and used resources of the Nanofabrication Laboratory and the Materials Characterization Laboratory at the Penn State Materials Research Institute in addition to the Advanced Light Source, which is a DOE Office of Science User Facility under contract no.\ DE-AC02-05CH11231. Special thanks to Chad Eichfeld and Michael Labella at the Penn State Materials Research Institute and Eric Gullikson at the Advanced Light Source.  

%





\bibliography{report}{}
\bibliographystyle{aasjournal}



\end{document}